\documentclass[aps,prb,twocolumn,showpacs,superscriptaddress]{revtex4}
\usepackage{amsmath,amssymb}
\usepackage{graphicx}
\usepackage{epsfig}
\usepackage[figuresright]{rotating}

\def\be{\begin{equation}} \def\ee{\end{equation}}
\def\bea{\begin{eqnarray}} \def\eea{\end{eqnarray}}

\def\nn{\nonumber}
\def\pp{\parallel}

\begin{document}

\title{Anisotropic Fermi liquid theory of the ultra cold fermionic
polar molecules: Landau parameters and collective modes}

\author{Ching-Kit Chan}
\affiliation{Department of Physics, University of California,
San Diego, CA 92093}
\author{Congjun Wu}
\affiliation{Department of Physics, University of California,
San Diego, CA 92093}
\author{Wei-Cheng Lee}
\affiliation{Department of Physics, University of California,
San Diego, CA 92093}
\author{S. Das Sarma}
\affiliation{Condensed Matter Theory Center, Department of Physics,
University of Maryland, College Park, MD 20742}

\begin{abstract}
We study the Fermi liquid properties of the cold atomic dipolar
Fermi gases with the explicit dipolar anisotropy using perturbative approaches.
Due to the explicit dipolar anisotropy, Fermi surfaces
exhibit distortions of the $d_{r^2-3z^2}$-type in three
dimensions and of the $d_{x^2-y^2}$-type in two dimensions.
The fermion self-energy, effective mass, and Fermi velocity develop 
the same anisotropy at the Hartree-Fock level proportional to the
interaction strength.
The Landau interaction parameters in the isotropic Fermi liquids
become the tri-diagonal Landau interaction matrices
in the dipolar Fermi liquids which renormalize thermodynamic susceptibilities.
With large dipolar interaction strength, the Fermi surface collapses 
along directions perpendicular to the dipole orientation.
The dynamic collective zero sound modes exhibit an anisotropic dispersion
with the largest sound velocity propagating along the polar directions.
Similarly,  the longitudinal $p$-wave channel spin mode becomes a 
propagating mode with an anisotropic dispersion 
in multi-component dipolar systems.
\end{abstract}
\maketitle

\section{{Introduction}\label{sect:introduction}}
The Fermi liquid theory is one of the most important paradigms
in contemporary condensed matter physics, which sets up the fundamental
framework to understand interacting fermion systems
\cite{leggett1975,baym1991}.
The most prominent feature of the Fermi liquid theory is the existence
of the Fermi surface and the long-lived low energy fermionic quasi-particles
around the Fermi surface.
The interaction effects, which may not be weak,
can be conveniently described by a set of
phenomenological Landau interaction parameters $F_l$ in different
partial wave channels characterized by the orbital angular momentum
number $l$.
Physical susceptibilities, including compressibility, specific heat,
and spin susceptibility receive important but still finite
renormalizations by the Landau interactions.
Furthermore, Fermi liquid states possess collective excitations
such as the zero sound mode whose restoring force is provided
by Landau interactions.

When Landau interaction parameters are negative and large enough, Fermi
liquid states may become unstable toward Fermi surface distortions named
the Pomeranchuk instabilities \cite{pomeranchuk1959}.
The most familiar example is ferromagnetism lying in the $s$-wave spin channel.
High partial wave channel instabilities in both density
and spin channels have been studied extensively in recent years
\cite{oganesyan2001,halboth2000,Nilsson2006,kee2003,wu2004,wu2007,varma2005,
hirsch1990}.
Another well-known, but of little practical
relevance, is the Kohn-Luttinger superconductivity driven by interactions
in the high angular momentum channels
\cite{kohn1965}.
The density Pomeranchuk instabilities in high partial channels
exhibit anisotropic Fermi surface distortions which are electronic
version of the nematic liquid crystal states
\cite{oganesyan2001,halboth2000,Nilsson2006,kee2003}.
Non-$s$-wave spin channel Pomeranchuk instabilities are essentially
``unconventional magnetism'', in analogy to unconventional
superconductivity.
They include both isotropic and
anisotropic phases dubbed $\alpha$ and $\beta$-phases as the counterparts
of $^3$He-$B$ (isotropic) and $A$ (anisotropic) phases
\cite{wu2004,wu2007,varma2005,hirsch1990,oganesyan2001},
respectively \cite{leggett1975}.

On the other hand, the rapid experimental progress of cold atomic
physics provides an exciting opportunity to study quantum many-body
systems with electric and magnetic dipolar interactions
\cite{ospelkaus2008,ni2008,griesmaier2005,mcclelland2006,lu2009}.
When electric dipole moments are aligned by the external field,
dipolar interaction decays as $1/r^3$, and thus is long-ranged 
in three dimensions (3D) but remains short-ranged in two dimensions (2D).
More importantly, the most prominent feature of dipolar interaction
is its spatial anisotropy  which possess the $d_{z^2-3r^2}$ anisotropy in 3D
and the $d_{x^2-y^2}$ anisotropy in 2D, respectively.
Considerable progress has been made in studying exotic properties
in anisotropic condensations of the dipolar bosonic systems as
reviewed in Refs \cite{koch2008,lahaye2009,lahaye2009a,menotti2007a}.

Furthermore, dipolar fermionic systems provide another exciting
opportunity to study exotic anisotropic many-body physics of fermions,
including anisotropic Fermi liquid states and Cooper paring states.
Physical observable should exhibit the same anisotropy accordingly
such as the shape of the Fermi surface \cite{sogo2008,miyakawa2008,
chang2009}.
Anisotropic Cooper pairing and Wigner crystallization
of dipolar Fermi gases has been theoretically investigated
\cite{baranov2002,baranov2004,baranov2008a,baranov2008b,bruun2008}.
Recently, Fregoso {\it et al.} \cite{fregoso2009} studied
the biaxial nematic instability in the dipolar Fermi gases as 
the $d$-wave channel Pomeranchuk instability, and generalized
the Landau interaction parameters to the tridiagonal Landau
interaction matrices.

Anticipating a great deal of experimental and theoretical activity in the
near future in polar molecular and atomic interacting fermionic systems,
we have provided in this article a comprehensive Landau Fermi liquid
theory for polar interactions including the full effects of anisotropy. 
The explicit dimensionless perturbation parameter is the
ratio between the characteristic dipolar interaction energy
and the Fermi energy.
The standard textbook isotropic Fermi liquid theory is generalized
into the anisotropic version which exhibits many different features
of both single-fermion and collective properties.
Our theory is a leading-order perturbative theory in the polar interaction
coupling constant, which is equivalent to a Hartree-Fock approximation of
the interaction.  We expect our theory to be quantitatively accurate in
the weak coupling regime, but the qualitative aspects of our theory, e.g.
the effect of anisotropy on the Fermi liquid parameters, should be
generally valid.
Our work only considers the explicit anisotropy due to the alignment
of the molecular electric dipolar moment.
The spontaneous anisotropic phase as a ferro-nematic state of the
dilute magnetic dipolar Fermi gases has been recently studied
by Fregoso {\it et al.} in Ref. \cite{fregoso2009a}

In Sect. \ref{sect:interaction}, the Fourier transforms of the
dipolar interactions in 3D and 2D are summarized, which nicely exhibit
the $d_{r^2-3z^2}$ symmetry and the $d_{x^2-y^2}$ symmetry, respectively.
The anisotropic interactions result in anisotropic Fermi surfaces.
The angular dependence of Fermi wavevectors, Fermi velocities,
and effective masses are presented in Sect. \ref{sect:HF}.
All of them exhibit the same isotropy proportional to the
the dipolar interaction strength at the leading order.

Because the dipolar interaction mixes different partial wave channels,
the Landau interaction parameters in isotropic systems becomes
Landau interaction matrices.
In Sect. \ref{sect:landau}, the Landau interaction matrices in 3D
calculated in Ref. \cite{fregoso2009}
are reviewed, and those in 2D are constructed
Fortunately due to the $d_{z^2-3r^2}$-symmetry of the dipolar interaction,
the mixing turns out to result in the tri-diagonal matrix.
between $l$ and $l\pm 2$ in 3D.
The thermodynamic susceptibilities would be renormalized by the
generalized Landau matrices as will be discussed in Sect. \ref{sect:thermo}.
In this section, we will also explore the thermodynamic instability, where Fermi surface
collapses along directions perpendicular to the dipole orientation.

In Sect. \ref{sect:collective mode}, we study the 3D collective modes in
both the density and spin channels, respectively,
focusing on the anisotropy effect.
The dynamic collective zero sound mode exhibits an anisotropic dispersion.
The sound velocity is largest if the propagation direction is along the
north and south poles, and becomes softened as the propagation deviates
from them.
For directions close to the equator of the Fermi surface, the zero
sound cannot propagate.
The spin-channel collective modes do not exist in the $s$-wave spin
wave channel.
Nevertheless, well-defined propagating collective modes appears in
the longitudinal $p$-wave spin channel, which has not been observed
in condensed matter systems before.

Conclusions and outlooks are made in Sect. \ref{Conclusions}.

\section{{The dipolar interaction in three and two dimensions}}
\label{sect:interaction}

The most prominent feature of the dipolar interaction is its spatial
anisotropy, {\it i.e.}, it depends on not only the distance
between two dipoles but also the angles between their relative
vectors and dipole orientations. In the case that all the dipoles
are orientated along the external electric field $\vec E$ set as
the $z$-axis, the dipolar interaction between dipoles at $\vec
r_{1}$ and $\vec r_2$ reads
\bea V_{3D}(\vec r_1-\vec r_2)&=&{d^2
\over {|\vec r_1-\vec r_2|^3}}
(1-3 \cos^2 \theta_{\vec r_1-\vec r_2} ) \nn \\
&=& -{2 d^2 \over {|\vec r_1-\vec r_2|^3}} P_2 (\cos \theta _{\vec
r_1-\vec r_2}), \label{eq:interaction_3D} \eea
where $\theta_{\vec r_1-\vec r_2}$ is the angle between $\vec r_1 -\vec r_2$
and the $\vec E$-field; $d$ is the electric dipole moment. The anisotropy
exhibits in the angular dependence of the $V_{3D}$ with the form
of the second Legendre polynomial.
$V_{3D}$ is repulsive for $\bar \theta_0
<\theta_{\vec r_1-\vec r_2} < \pi-\bar \theta_0$, and
attractive otherwise, where
\bea
\bar\theta_0=\cos^{-1}\frac{1}{\sqrt 3}\approx 55^\circ.
\label{eq:thetabar}
\eea

\begin{figure}[tb]
\centering\epsfig{file=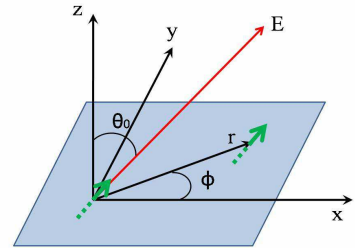,clip=1,
width=0.6\linewidth,angle=0}
\caption{Schematic sketch of a 2D dipolar system. The external
electric field $\vec{E}$ controls the dipole orientation and
thus the dipolar interaction.}
\label{fig:interaction_2D}
\end{figure}

If the spatial locations of dipoles are confined in a two dimensional
plane, and the $\vec E$-field is set in the $xz$-plane with an
angle of $\theta_0$ relative to the plane as depicted in
Fig. \ref{fig:interaction_2D}, then Eq. \ref{eq:interaction_3D} reduces to
\bea
&&V_{2D}(\vec{r_1}-\vec r_2,\theta_0)={d^2 \over {|\vec r_1-\vec r_2|^3}}
(1-3 \sin^2 \theta_0 \cos^2 \phi) \nn \\
&&={d^2 \over {|\vec r_1-\vec r_2|^3}}
\big\{P_2(\cos\theta_0)-\frac{3}{2} \sin^2 \theta_0 \cos 2\phi
\big\}, \label{eq:interaction_2D} \eea
 where $\phi$ is the
azimuthal angle relative to the $x$-axis. Eq.
\ref{eq:interaction_2D} is decomposed into the isotropic component
and an anisotropic $d$-wave component, whose relative weight is
tunable by varying the parameter angle $\theta_0$. If $\vec E$ is
perpendicular to the plane, {\it i.e.} $\theta_0=0$, $V_{2D}$ is
isotropic and repulsive. On the other hand, at $\theta_0 =
\bar{\theta}_0$, $V_{2D}$ is purely
anisotropic with the $d$-wave form factor $\cos 2\phi$. As the
external electric field is tilted, {\it i.e.}, $\theta_0$ varies
from $0$ to $\frac{\pi}{2}$, the 2D dipolar interaction gradually
changes from an isotropic repulsive to an attractive one at large
value of $\theta_0$.

The anisotropies in the Fourier transform of dipolar interactions in
both 3D and 2D exhibit in their dependencies on momentum orientations.
In order to handle
both the long range part and the short distance divergence of the
dipolar interactions, we introduce a small distance cutoff
$\epsilon$ beyond which the dipolar interactions Eq.
\ref{eq:interaction_3D} and Eq. \ref{eq:interaction_2D} are valid,
and a large distance cutoff $R$. $\epsilon$ can be chosen, say,
one order larger than the size of the dipolar molecule and $R$ is
the radius of the system.

\subsection{Fourier transform of the 3D dipolar interaction}

For the 3D dipolar interaction,
its Fourier transform can be performed as
 \bea V_{3D}(\vec
q)&=&8\pi d^2 \big\{\frac{j_1(q\epsilon)}{q\epsilon} -\frac{j_1(q
R)}
{q R}\big\} P_2(\cos\theta_{\vec q}) \nn \\
&\longrightarrow& \frac{8\pi d^2}{3} P_2(\cos \theta_{\vec q}),
\label{eq:interaction_3Dk}
\eea
at $q\epsilon\rightarrow 0$ and $q R\rightarrow \infty$.
Its angular dependence is the
second Legendre polynomial of momentum direction inherited from
the real space form of Eq. \ref{eq:interaction_3D}; $j_1(x)$ is
the first order spherical Bessel function with the asymptotic
behavior
$j_1(x) \rightarrow \frac{x}{3}$ as
$x\rightarrow 0$; and
$j_1(x) \rightarrow  \frac{1}{x} \sin
( x- \frac{\pi}{2})$ as  $x\rightarrow \infty$.

Eq. \ref{eq:interaction_3Dk} does not depend on the magnitude of $q$
after the limits of $q\epsilon\rightarrow 0$ and $q R\rightarrow +\infty$
are taken because of the spatial integral over $1/r^3$ renders the
result dimensionless.
At $\vec q=0$, $V_{3D}(\vec q=0)=0$ arising from the fact that
the spatial average of the 3D dipolar interaction is zero.
However, $V_{3D} (\vec q)$ is not analytic as $\vec q\rightarrow 0$
in the limit of $R\rightarrow +\infty$ due to the angular variation.
In fact, the smallest value of $qR$ is at the order of $O(1)$,
this non-analyticity even exist for large but finite
value of $R$.

An interesting feature of the above Fourier transforms is that $V_{3D}
(\vec q)$ is  most positive when $\vec q$ is along the $z$-axis, but
most negative when $\vec q$ is in the equator plane, which is
just opposite to the case of $V_{3D}(\vec r)$ in real space.
This can be intuitively understood as follows.
Considering a dipole density wave propagating along the $z$-axis,
the wave fronts (equal phase lines) are perpendicular to the dipole
orientation, thus the interaction energy is repulsive.
On the other hand, if the dipole density wavevector lies in the equator,
then the wave fronts are parallel to the dipole orientation which
renders the interaction negative.

\subsection{Fourier transform of the 2D dipolar interaction}
\label{subsect:2Dfourier}

In two dimensions, the Fourier transform of Eq. \ref{eq:interaction_2D}
is more subtle, which can be expressed as
\bea
V_{2D}(\vec{q},\theta_0)&=& 2\pi d^2 P_2(\cos \theta_0)
\Big \{\frac{1}{\epsilon} - \frac{J_0(qR)}{R}\nn \\
&+& q \big [J_1(q\epsilon)- J_1(qR)\big ] -q I_2(q\epsilon,qR)
\Big \}
 \nonumber \\
&+& \pi d^2 \sin^2\theta_0 \cos 2 \phi_q ~q \Big\{
3 \big [\frac{J_2(qR)}{qR}-\frac{J_2(q\epsilon)}{q\epsilon}\big ] \nn \\
&-& \big [ J_3(qR)-J_3(q\epsilon) \big ]+I_2(q\epsilon, qR) \Big\},
\label{eq:interaction_2Dk}
\eea
where $J_{0,1,2}(x)$ are the Bessel functions of the 1st, 2nd, and 3rd
orders, respectively;
$I_2$ is the integral defined as
\bea
I_2(q\epsilon, qR)=\int^{qR}_{q\epsilon} dx J_2(x),
\eea
and $I_2(0,+\infty)=1$.
In the regular limit of $q\epsilon\rightarrow 0$ and
$qR\rightarrow +\infty$, the complicated form of Eq.
\ref{eq:interaction_2Dk} can be simplified into
\bea
V_{2D}(\vec{q},\theta_0)&=& 2\pi d^2 P_2(\cos \theta_0)
(\frac{1}{\epsilon} -q)\nn \\
&+& \pi d^2 \sin^2\theta_0 ~q \cos 2 \phi_q.
\label{eq:fourier_2D}
\eea
In particular, at $\theta_0=\bar \theta_0$, Eq.
\ref{eq:fourier_2D} is entirely anisotropic,
\bea
V_{2D}(\vec{q},\theta_0=\bar{\theta}_0)={2 \pi d^2\over {3}} q
\cos 2 \phi_q,
\label{eq:interaction_2D_k_theta_0}
\eea
as one can see from the real space interaction in
Eq. \ref{eq:interaction_2D}. Eq. \ref{eq:fourier_2D} even holds
in the long wavelength limit $q R \rightarrow O(1)$.
On the other hand, for the short wavelength limit $x=q\epsilon$
at the order of $O(1)$, additional terms should be added into Eq.
\ref{eq:fourier_2D} as \bea \Delta V_{2D}(\vec q\rightarrow 0,
\theta_0) &=&2\pi d^2 P_2(\cos \theta_0) \frac{1}{\epsilon}
[x J_1(x) + x I_2(0, x)] \nn \\
 &+&\pi d^2 \sin^2\theta_0 \cos \phi_q \frac{1}{\epsilon}
[ 3 J_2(x)+x J_3(x)\nn \\
&-&x I_2(0,x)]. \eea
In the dilute limit, where the Fermi wavevector $k_f$ satisfies
$k_f \epsilon \ll 1$, we only need to use $q$ up to the order
of $k_f$.
Thus in the sections below, we shall use Eq. \ref{eq:fourier_2D}
as the 2D Fourier transform of our dipolar interaction.

The anisotropic component of $V_{2D}(\vec q)$ (second term of 
Eq. \ref{eq:fourier_2D}) has a similar feature to the 3D case,
which is repulsive along the $x$-axis but negative along $y$-axis.
Notice that there is 
also an isotropic component for $V_{2D}$ which can be either
positive or negative depending on the external parameter $\theta_0$.
Therefore, in momentum space, the dipolar interaction in 2D contains
both $s$ and $d$-wave components, while it only has a $d$-wave symmetry in 3D.

\begin{figure}[tb]
\centering\epsfig{file=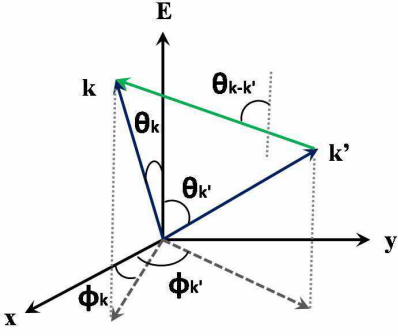,clip=1,width=0.85\linewidth,angle=0}
\caption{
The dipole moments polarized by the external $\vec E$-field are along the
$z$-axis. The polar and azimuthal angles of $\vec k$ and $\vec k^\prime$
are denoted by $\theta_{k(k^\prime)}$ and $\phi_{k(k^\prime)}$, respectively.
$\theta_{\vec k-\vec k^\prime}$ is the polar angle of the vector difference
of $\vec k-\vec k^\prime$ but not the angle between $\vec k$
and $\vec k^\prime$.}
\label{fig:3d_vectors}
\end{figure}

\section{Hartree-Fock Self-energy and Fermi Surface Deformation}
\label{sect:HF}

The anisotropy of the dipolar interaction exhibits in the fermion
self-energy at the Hartree-Fock level, which results in anisotropic
Fermi surface distortions as studied in Refs \cite{fregoso2009, sogo2008,
miyakawa2008}.
A convenient dimensionless parameter to describe the interaction strength
is defined in 3D and 2D as \cite{fregoso2009}
\bea
\lambda_{3D}&=&\frac{E_{int}^{3D}}{E^{3D}_{k_{f0}}}=\frac{d^2 m k_{f_0}^{3D}}{3\pi^2 \hbar^2}, \nn \\
\lambda_{2D}&=&\frac{E_{int}^{3D}}{E^{2D}_{k_{f0}}}=\frac{d^2 m k_{f_0}^{2D}}{4 \pi^{3/2}
\hbar^2},
\label{eq:strength}
\eea
where $k_{f0}^{3D,2D}$ are the non-interacting Fermi wavevectors in 3D and
2D, respectively;
$E_{int}$ is the average dipolar interaction and
$E_{k_{f0}}^{3D,2D}=\frac{\hbar^2 (k^{3D,2D}_{f0})^2}{2m}$ are
the kinetic energies at the Fermi surface in the absence of interaction.
Since that $\lambda_{3D}$ and $\lambda_{2D}$ are at the same order,
we will not distinguish them but use $\lambda=\lambda_{3D}$ for both 2D and 3D.
In this section, we shall evaluate the Hartree-Fock self-energy in 2D and 3D
perturbatively at the linear order of $\lambda$  from which the Fermi
surface distortion will be determined.
We will see that both the Hartree-Fock self-energies and Fermi surface distortions
possess the same symmetries inherited from the corresponding interactions.

\subsection{Anisotropic Hartree-Fock self-energy in 3D}

\begin{figure}[tb]
\centering\epsfig{file=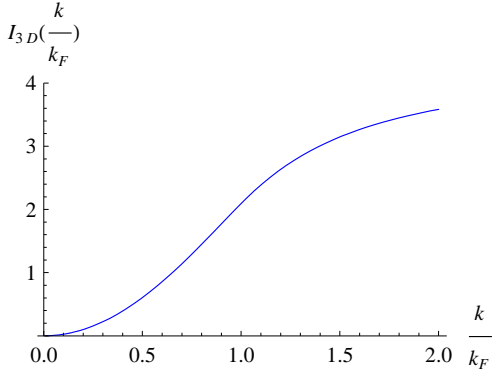,clip=1,width=0.75\linewidth,angle=0}
\caption{The dimensionless $k$-dependent function
$I_{3D}(\frac{k}{k_{f}})$ of the 3D Hartree-Fock self-energy in Eq. \ref{eq:HF_3D}. }
\label{fig:HF3D}
\end{figure}

We first consider the situation in 3D, the Hartree-Fock
self-energy can be expressed as
\begin{eqnarray}
\Sigma^{HF}_{3D} (\vec{k})&=&\frac{1}{V}
\sum_{k'} \big\{ V_{3D}(\vec q=0)-V_{3D}(\vec{k}-\vec{k'}) \big\}
n_{\vec{k'}} \nn \\
&=& -\frac{8\pi d^2}{3 V}\sum_{k'} P_2(\cos \theta_{\vec k-\vec k^\prime})
n_{\vec k^\prime}\nn \\
&=& -\frac{8\pi d^2}{3} \int_{k^\prime<k_{f0}} \frac{d^3 k^\prime}{(2\pi)^3}
P_2(\cos \theta_{\vec k-\vec k^\prime})
\label{eq:HF}
\end{eqnarray}
where $n_{\vec k}$ is the Fermi occupation number; the Hartree
contribution goes to zero. In Eq. \ref{eq:HF}, the leading
contribution of the anisotropy to $\Sigma^{HF}(\vec k)$ comes from
the interaction.
The effect from the Fermi surface distortion is at a higher order
of $\lambda$, and thus is neglected, thus we will take the Fermi
wavevector in Eq. \ref{eq:HF} as that of zero dipolar interaction
$k_{f0}^{3D}$.
$\theta_{\vec{k}-\vec{k'}}$ is the angle between the momentum transfer
$\vec{k}-\vec{k'}$ and $\vec E$, which satisfies
\bea
\cos\theta_{\vec k-\vec k^\prime}&=&\frac{ (\vec k-\vec k^\prime)
\cdot \hat E} {|\vec k-\vec k^\prime|}
\eea
Please notice that $\theta_{\vec{k}-\vec{k'}}$ is {\it not} the angle
between $\vec{k}$ and $\vec{k'}$ (see Fig. \ref{fig:3d_vectors}).

This Hartree-Fock self-energy can be evaluated analytically as
\begin{eqnarray}
&&\Sigma^{HF}_{3D}(\vec{k})=-
2\lambda E_{k_{f0}}^{3D} P_2(\cos \theta_k)\
I_{3D} ({k \over k_{f_0}^{3D}}),
\label{eq:HF_3D}
\end{eqnarray}
where $E_{k_{f0}}=\hbar^2 k_{f0}^2/2m$ and
\bea
I_{3D}(x)&= & \frac{\pi}{12}\Big \{ 3x^2+ 8
-{3\over x^2}
+\frac{3(1-x^2)^3}{2x^3}
\ln|\frac{1+x}{1-x}| \Big \} \nn \\
\eea is a
monotonically increasing function as depicted in Fig. \ref{fig:HF3D}.
This form agrees with the self-energy calculated for the
3D dipolar Fermi gases in Ref. \cite{fregoso2009}.
Right on the Fermi surface,
\bea
\Sigma_{3D}^{HF}(k=k_{f0}^{3D})
&=&-\frac{4\pi}{3}\lambda E_{k_{f0}}^{3D}P_2(\cos\theta_k).
\eea
Naturally Eq. \ref{eq:HF_3D} exhibits the $d$-wave form factor $P_2(\cos
\theta_k)$.
We shall see more examples below where the physical
quantities possess the symmetry originated from the dipolar
nature.


\subsection{Anisotropic Hartree-Fock self-energy in 2D}

\begin{figure}[tb]
\centering\epsfig{file=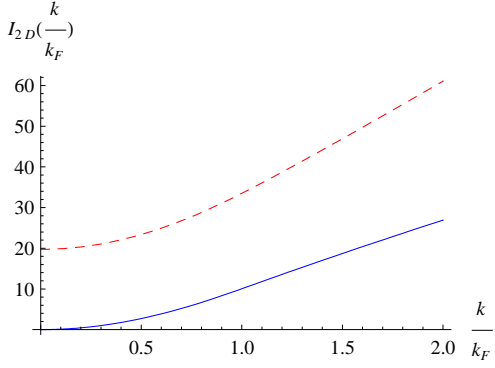,clip=1,width=0.75\linewidth,angle=0}
\caption{The $k$-dependent components of the 2D HF self-energies in 2D.
The solid (blue) and dashed (red) curves represent the
dimensionless functions $I_{2D}^{ani}(k/k_{f})$ and
$I_{2D}^{iso}(k/k_{f})$ defined in Eq. \ref{eq:HF_2D}.
}
\label{fig:HF2D}
\end{figure}

Similarly, for 2D system, the HF self-energy is evaluated as
\begin{eqnarray}
\Sigma^{HF}_{2D}(\vec{k},\theta_0)&=&  \lambda E_{k_{f0}}^{2D}
\Big\{ P_2(\cos\theta_0)\ I_{2D}^{iso} \Big( \frac{k}{k_{f0}^{2D}} \Big) \nn \\
&-& \sin^2\theta_0 \cos 2\phi_k\ I_{2D}^{ani} \Big ({k \over k_{f0}^{2D}}\Big)
\Big \},
\label{eq:HF_2D}
\end{eqnarray}
where $k_{f0}^{2D}$ is the Fermi wavevector at zero dipolar interaction and
\bea
I_{2D}^{iso}(x)&=& 12\pi \int_0^1dx' x' |x+x'| E\Big(\frac{2\sqrt{xx'}}{x+x'}\Big) \nn \\
I_{2D}^{ani}(x)&=& 2 \pi \int_0^1dx' \frac{x'|x+x'|}{x^2} \Big\{(x-x')^2 K\Big(\frac{2\sqrt{xx'}}{x+x'}\Big) \nn \\
&& \ \ \ \ \ \ \ \ \ \ \ +\ \ (2x^2-x'^2) E\Big(\frac{2\sqrt{xx'}}{x+x'}\Big) \Big\}.
\eea
The functions $K(y)$ and $E(y)$ are the standard complete elliptic
integral of the first and second kinds, respectively as
\bea
K(y)&=& \int^{\frac{\pi}{2}}_0
d\alpha \frac{1}{\sqrt { 1- y^2 \sin^2 \alpha} },
\nn \\
E(y)&=& \int^{\frac{\pi}{2}}_0 d\alpha \sqrt { 1- y^2 \sin^2 \alpha}.
\eea
The behavior of $I_{2D}^{iso}(x)$ and $I_{2D}^{ani}(x)$ are plotted
in Fig. \ref{fig:HF2D}. Eq. \ref{eq:HF_2D} indicates that the self-energy
in 2D comprises of an isotropic and an anisotropic terms.
The former shifts the chemical potential $\mu$, while the latter
distorts the Fermi surface.
At $k=k_{f0}^{2D}$,  Eq. \ref{eq:HF_2D} reduces into
\bea
&&\Sigma_{2D}^{HF}(k=k_{f0}^{2D}, \phi_k,\theta_0) \nn \\
&=& -\lambda E_{k_{f0}}^{2D} \{-\frac{32\pi}{3} P_2(\cos\theta_0) +\frac{16\pi}{5} \sin^{2}\theta_0 \cos2\phi_k\}.\nn \\
\eea
Again, Eq. \ref{eq:HF_2D} contains the same angular dependence as
the 2D interaction does. Both Eq. \ref{eq:HF_3D} and \ref{eq:HF_2D}
indicate that the Hartree-Fock-self energy changes monotonically with the
momentum $k$ in 2D and 3D (Fig. \ref{fig:HF3D} and Fig. \ref{fig:HF2D}).

\subsection{Hartree-Fock self-energy for two-component dipolar systems}

In two-component dipolar systems, the exchange part of the Hartree-Fock
self-energy
only exists intra-component interaction, and the Hartree part exists
in both intra and inter-component interactions.
Since the Hartree contribution vanishes in the 3D dipolar systems,
the Hartree-Fock self-energy remains the same as in the single-component
systems.
In the 2D dipolar systems, there is an extra contribution
from the inter-component Hartree interaction as
\bea
\Delta \Sigma_{2D}^{HF}= \frac{3\pi^2}{k^{2D}_{f0}\epsilon} \lambda E_{k_{f0}} P_2(\cos \theta_0),
\eea
where $\epsilon$ is the 2D phenomenological cutoff constant we discussed before. This extra self-energy contribution is momentum independent and can be offset by an overall shift
of the chemical potential.

\subsection{Anisotropic Fermi surface distortions}

The anisotropic Hartree-Fock self-energy from the dipolar interaction naturally
results in anisotropic Fermi surface distortions.
With Eq. \ref{eq:HF_3D} and \ref{eq:HF_2D}, we determine the
distortions by solving:
\bea
\epsilon_{HF}(\vec k_f)=\epsilon_0(\vec
k_f)+\Sigma^{HF}(\vec k_f )=\mu(n,\lambda).
\eea
Here, $n$ is the particle density and we recall that $k^{3D}_{f_0}$ is the
Fermi wavevector for $\lambda=0$.
With nonzero $\lambda$, the dependence of $\vec k_f$ on the
polar angle in 3D is solved as
\bea
\frac{k_f^{3D}(\theta_k)}{k_{f0}^{3D}}=
1-\frac{4\pi^2}{45}\lambda^2  + \frac{2\pi}{3}\lambda P_2(\cos
\theta_k) \label{eq:fermi_k_3D}.
\eea
Similarly, in 2D, the dependence of $\vec k$ on the azimuthal angle is
solved as
\bea
\frac{k_f^{2D}(\phi_k,\theta_0)}{k_{f0}^{2D}}&=&
1-\frac{16\pi^2}{25} \sin^4\theta_0
\lambda^2 \nn \\
&+& \frac{8\pi}{5} \sin^2\theta_0 \lambda \cos 2\phi_k.
\label{eq:fermi_k_2D}
\eea

The anisotropic Fermi surface distortions in 2D and 3D show
linear dependence on dipolar interaction strength $\lambda$. Due
to particle number conservation, there are small shrinking of
Fermi momentum in both 2D and 3D at the quadratic order of
$\lambda$. Notice that these two equations are correct to the
order of $\lambda$. The $\lambda^2$ terms appear to conserve
particle numbers due to this lowest order Fermi surface
deformation. When higher order contribution is considered,
anisotropy can also enter in the second order of $\lambda$ and
particle conservation will in turn lead to an isotropic
$\lambda^3$ correction.

\begin{figure}[tb]
\centering\epsfig{file=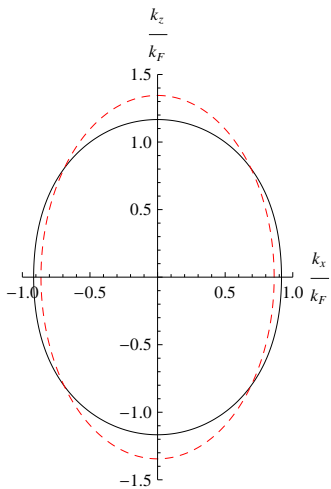,clip=1,
width=0.5\linewidth,angle=0}
\caption{The deformed Fermi surface for $\lambda=\frac{1}{2 \pi}$
in the 3D dipolar system obtained by the perturbative (solid)
and variational (dashed red) approaches.
The external electric field lies along the z axis here.}
\label{fig:fermi_surface_3D}
\end{figure}

\begin{figure}[tb]
\centering\epsfig{file=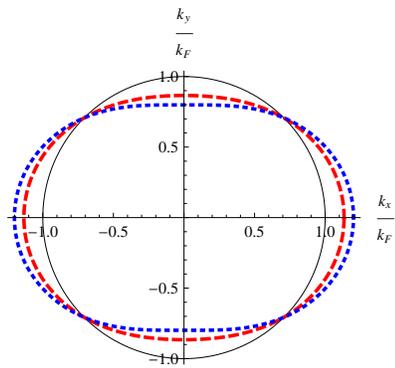,clip=1,
width=0.6\linewidth,angle=0}
\caption{The 2D distorted Fermi surfaces for $\lambda=0.02$. The solid (black), dashed (red) and blue (dotted) curves corresponds to an external electric field configuration at $\theta_0$=0, $\bar\theta_0$ and $\frac{\pi}{2}$ respectively. The electric field is projected to x-axis in this 2D system.}
\label{fig:fermi_surface_2D}
\end{figure}

We emphasize here again that the results obtained based on
Eq. \ref{eq:HF_3D} and Eq. \ref{eq:HF_2D} are perturbative
and are correct for $\lambda \ll 1$, while they provide
qualitative features for $\lambda < 1$.
As a comparison with the variational approach suggested
in Ref. \cite{miyakawa2008},
we plot the deformed Fermi surface in Fig. \ref{fig:fermi_surface_3D}
for $\lambda=\frac{1}{2\pi}$ in 3D,
which corresponds physically to the stability limit of the
dipolar system (to be discussed in Sect. \ref{sect:thermo}).
Our perturbation results give rise a less prolate shape
of the Fermi surface than those obtained from the variational
approach. Fig.\ref{fig:fermi_surface_2D} shows the same 
distorted Fermi surfaces in 2D. At $\theta_0=0$, the interaction 
is isotropic and does not deform the Fermi surface. 
It becomes prolate when $\theta_0$ increases. We would like to
emphasize that the interaction is most anisotropic 
at $\theta_0=\bar\theta_0$, though a larger $\theta_0$ 
will stretch the Fermi surface more.

\subsection{Anisotropic Fermi velocities and effective masses}
\label{sect:anisotropic_fermi_surface}

The Fermi velocities become anisotropic in both 2D and 3D
at non-zero $\lambda$.
In comparison to interacting electron systems, the Hartree-Fock contributions
to the 2D and 3D dipolar Fermi gases are less singular, {\it i.e.,},
the radial derivatives at the Fermi surface remain continuous:
the Hartree correction of Fermi velocity is continuous because
the Fourier transforms of the dipolar interaction are finite as $\vec
q\rightarrow 0$.
Furthermore, the difference of the Hartree-Fock self-energies for the single
component and multi-component dipolar gases is momentum independent,
thus the Fermi velocities are the same for both cases and
we will not distinguish them in this subsection.

We first consider the 3D case.
After taking into account the Hartree-Fock correction, the projections of
Fermi velocities along the radial
direction $\hat k$ and angular direction $\hat e_\theta$  are defined as
$\hat k \cdot \vec v_f^{3D} (\vec k)$ and $
\hat e_{\theta} \cdot \vec v_f^{3D} (\vec k)$
as
\bea
\hat k \cdot \vec
v_f^{3D}(\vec k)&=& \frac{k_{f}^{3D}(\theta_k)}{m}+ \frac{\partial \Sigma^{HF}_{3D}(k,
\theta_k)}{\partial k}, \nn \\
\hat e_{\theta} \cdot \vec v_f^{3D} (\theta_k)&=&
\frac{\partial \Sigma^{HF}_{3D}(k,\theta_k)}{ k_{f}^{3D}
\partial \theta_k}.
\eea
Taking into account the angular dependence of Fermi wavevector
in Eq. \ref{eq:fermi_k_3D} and \ref{eq:fermi_k_2D}, we arrive
at the radial and angular Fermi velocities
corrections to the linear order of $\lambda$ as
\bea
\hat k_f \cdot \vec v_f^{3D}(\theta_k)
&=& \frac{k_{f0}^{3D}}{m} \Big\{1 - \frac{\pi}{3} \lambda
P_2(\cos\theta_k) \Big\},  \nn \\
\hat e_{\theta} \cdot \vec v_f^{3D} (\theta_k)&=&
\frac{k_{f0}^{3D}}{m}\pi  \lambda \sin 2\theta_k.
\label{eq:fermivelocity_3D}
\eea
There are two opposite contributions to $\hat k_f \cdot \vec v_f^{3D}$:
(1) the Fermi surface deformation, $k_f(\theta_k)$ and
(2) the Hartree-Fock self-energy modification.
In 3D dipolar system, the latter is stronger and thus the radial Fermi velocity
is suppressed at north and south poles but enhanced along
the equator of the Fermi surface.

According to Eq. \ref{eq:fermivelocity_3D}, we define the longitudinal
and transverse effective masses $m^*_{3D,\pp}$ and $m^*_{3D,\perp}$ in 3D as
\bea
\label{eq:radmass}
\frac{1}{m^*_{3D,\pp}(\theta_k)}&=&\frac{\hat k_f \cdot \vec v_f^{3D}(\theta_k)}{
k^{3D}_{f}(\theta_k)} = \frac{1}{m} \Big\{ 1 -\pi\lambda
P_2(\cos \theta_k)\Big\}, \nn \\
\frac{1}{m^*_{3D,\perp}(\theta_k)}&=&\frac{\hat e_{\theta_k} 
\cdot \vec v_f^{3D}(\theta_k)}{
k^{3D}_f(\theta_k)}= \frac{1}{m}\pi\lambda
\sin 2\theta_k.
\eea

For the 2D case, the projections of Fermi velocity in 2D
along the radial $\hat k$ and the azimuthal angle
$\hat e_{\phi_k}$-directions are calculated as
\bea
\label{eq:2d_effectmass}
\hat k \cdot \vec v_f^{2D}(\phi_k,\theta_0) 
&=&\frac{k_{f0}^{2D}}{m} \Big\{1+ \lambda \Big[ 4 \pi P_2(\cos\theta_0) \nn \\ 
&-& \frac{6\pi}{5}\sin^2\theta_0 \cos2\phi_k \Big ]\Big\}, \\
\hat e_{\phi_k}\cdot \vec v_f^{2D}(\phi_k,\theta_0)&=&
\frac{k_{f0}^{2D}}{m}\frac{16\pi}{5}  \lambda\sin^2\theta_0\sin 2\phi_k.
\eea
For a general value of $\theta_0 \neq \bar\theta_0$, there is an
additional isotropic renormalization to $v_{f}^{2D}(\phi_k,\theta_0)$
in Eq. \ref{eq:2d_effectmass},
which comes from the isotropic part of the 2D dipolar interaction.

This means that in 2D, we also have a reduced radial Fermi velocity at
$\phi_k=0,\pi$, while it is boosted when $\phi_k=\pm \frac{\pi}{2}$.
Accordingly, similar to the 3D situation, we can also compute the longitudinal
and transverse effective masses $m^*_{2D,\pp}$ and $m^*_{2D,\perp}$ in 2D as
\bea
&&\frac{1}{m^*_{2D,\pp}(\phi_k,\theta_0)} \nn \\
&&\ \ \ \ =\frac{1}{m} \Big\{ 1+\lambda \Big [4\pi P_2(\cos\theta_0) 
-\frac{14\pi}{5}
\sin^2\theta_0\cos 2\phi_k\Big ] \Big\}, \nonumber \\
&&\frac{1}{m^*_{2D,\perp}(\phi_k,\theta_0)}=\frac{ 1}{m} \frac{16\pi}{5}\lambda\sin^2\theta_0 
 \sin 2\phi_k.
\eea

\subsection{Renormalization of density of states}
Now we study the renormalization of the density of states (DOS)
at the Fermi surface due to the dipolar interaction at the Hartree-Fock level.
We define the 3D differential DOS of the single component
$N^{3D}(\Omega_k)$ in the direction of $\Omega_k$ as
\bea
N^{3D}(\Omega_k)\frac{d\Omega_k}{4\pi}&=&
 \frac{k^{3D}_{f}(\theta_k) d \Omega_k}{(2\pi)^3 v_f^{3D}(\theta_k)}
\sqrt{[k^{3D}_f(\theta_k)]^2+(\frac{d k_f^{3D}}{d\theta_k})^2} 
\nn \\
&=&\frac{m k_{f0}^{3D}}{\hbar (2\pi)^3}
[1+\frac{5\pi}{3}\lambda P_2(\cos\theta_k)] d\Omega_k.
\label{eq:3DDOS}
\eea

At the linear order of $\lambda$, $N^{3D}(\Omega_k)$
develops the same anisotropy of $P_2(\cos\theta)$.
The total DOS reads
\bea
N^{3D}=\int \frac{d\Omega_k}{4\pi} N^{3D}(\Omega_k),
\eea
which does not change at the linear order of $\lambda$ compared
to that of the free Fermi gas.
This means that the specific heat, which is proportional to
the total DOS at the Fermi surface, is not renormalized to the
linear order of $\lambda$.
In considering the actual area of Fermi surface is enlarged, the correction
at higher orders of $\lambda$ should increase the total DOS.

Similarly, we consider the anisotropic case in 2D.
The 2D differential DOS of the single component $N^{2D}(\Omega)$
at the Fermi surface along the direction of the azimuthal angle $\phi_k$
is defined as
\begin{widetext}
\bea
N^{2D}(\phi_k) \frac{d\phi_k}{2\pi}
= \frac{d\phi_k}{(2\pi)^2 v_f^{2D}(\phi_k)}
\sqrt{[k^{2D}_f(\phi_k,\theta_0)]^2+(\frac{d k_f^{2D}}{d\phi_k})^2} 
= \frac{m d\phi_k}{(2\pi)^2} \Big\{
1+\lambda \Big[ -4 \pi P_2(\cos\theta_0) 
+ \frac{14\pi}{5}\sin^2\theta_0  \cos2\phi_k \Big ]\Big\},
\nn \\
\eea
\end{widetext}
which clearly exhibits the $d$-wave anisotropy.
The integrated total DOS at 2D is not changed at the linear order
of $\lambda$ either.
Similarly, the 2D specific heat is not renormalized to the
linear order of $\lambda$.

\section{Landau Interactions for the dipolar Fermi liquids}
\label{sect:landau}
In this section, we construct the Landau Fermi liquid Hamiltonian
for the dipolar fermion gas which has also been studied by Fregoso
{\it et. al} \cite{fregoso2009} before.
In the isotropic systems, the interaction effects in the Fermi liquid
theory are captured by a set of Landau parameters $F_l$ in different
partial wave channels.
In dipolar systems, the anisotropic interaction leads to the mixing
of interactions in different partial wave channels, thus we need
to generalize the concept of Landau parameters into the Landau matrices.

\subsection{Landau interaction matrix for the single component dipolar
gases}

In order to have a common reference Fermi surface for different
but small values of $\lambda$, we choose it as that of the
free fermion gas with the same particle density.
Define the variation of the Fermi distribution at momentum $\vec k$,
\bea
\delta n_{\vec k}=n_{\vec k}-n_{0, k},
\label{eq:FSvariation}
\eea
where $n_0(k)=1-\theta(k-k_{f_0}^{3D})$.
Assuming a small variation of the fermion distribution $\delta n_{\vec k}$
close to the Fermi surface, the ground state energy of the Fermi liquid
state changes as
\bea
\delta E=\sum_k \epsilon_k \delta n_k +\frac{1}{2V}\sum_{\vec k, \vec k^\prime}
f(\vec k,\vec k^\prime)\delta n_{\vec k} \delta n_{\vec k^\prime},
\eea
where $\vec k, \vec k^\prime$ are momenta close to the Fermi surface;
$f(\vec k, \vec k^\prime)$ is the interaction function describing
the forward scattering amplitude.
$\epsilon_k$ includes the bare
parabolic dispersion $\epsilon_k^0$ and the self-energy correction
$\Sigma_{HF}(\vec k)$ which exhibits the  $d_{r^2-3z^2}$ anisotropy.
At the Hartree-Fock level, $f(\vec k, \vec k^\prime)$ is expressed as
$f(\vec k, \vec k^\prime)=V(\vec q =0)-V(\vec k-\vec k^\prime)$
where the first and second terms are the Hartree and Fock contributions,
respectively.
Due to the explicit rotational symmetry breaking,
$f(\vec k, \vec k^\prime)$ depends on directions of both $\vec k$
and $\vec k^\prime$, not just the relative angle between $\vec k$ and
$\vec k^\prime$ as in isotropic Fermi liquids.
For the 3D dipolar system, the classic Hartree term vanishes because
the spatial average of the dipolar interaction is zero.
However, due to the non-analyticity of the $V_{3D}(\vec q)$ as
$\vec q\rightarrow 0$, in the calculation of the zero sound
collective excitation in Sect. \ref{sect:collective mode},
the dependence of $\vec q$ in the Landau interaction needs to be
included as
\bea
\frac{1}{2V}\sum_{\vec k, \vec k^\prime} ~   f(\vec k, \vec k^\prime; \vec q)
~n_{\vec k, \vec q} ~n_{\vec k, -\vec q}.
\label{eq:landauHF}
\eea
where
$f(\vec k,\vec k';\vec q)=V(\vec q)- V(\vec k-\vec k^\prime)$
and $n_{\vec k, \vec q}=c^\dagger_{\vec k+\vec q} c_{\vec k}$.

\subsubsection{3D Landau interactions}
In this part, we review the Landau parameter calculation performed
in Ref. \cite{fregoso2009}.
Since quasi-particle excitations are close to the Fermi surface, we
integrate out radial direction and obtain the angular distribution.
In the 3D system,
\bea
\delta n (\Omega_{\vec k})=\int \frac{k^2 d k}{(2\pi)^3} \delta n_{\vec k}.
\eea
We expand $\delta n(\Omega_{\vec k})$ in terms of the spherical harmonics as
\bea
\delta n(\Omega_{\vec k})=\sum_{lm} Y_{lm} (\Omega_{\vec k})
\delta n_{lm},
\label{eq:n_lm}
\eea
where $Y_{lm}$'s satisfy the normalization convention
$\int d\Omega ~Y^*_{lm} (\Omega) Y_{lm} (\Omega) =1$.
Due to the anisotropy of the dipolar interaction,
its spherical harmonics decomposition becomes
\bea
f^{3D}(\vec k, \vec k^\prime)=\sum_{l,l^\prime;m}
\frac{4\pi f_{ll^\prime;m}^{3D} }{\sqrt {(2l^+1) (2l^\prime+1)} }
Y_{lm}^*(\Omega_k) Y_{l^\prime m} (\Omega_{\vec k^\prime}), \nn \\
\label{eq:landauhmncs}
\eea
where $f_{ll^\prime;m}$ remains diagonal for $m$ but couples partial wave
channels with $l^\prime=l, l\pm 2$.
This is a direct result from Wigner-Eckart theorem because the dipolar
interaction possesses the symmetry of  $d_{r^2-3z^2}$.
The mixing between $l^\prime$ and $l$ with $l^\prime=l\pm 1$ is forbidden
because the dipolar interaction is parity even.

For the 3D systems, the matrix elements $f_{ll^\prime;m}$ are tridiagonal and
have been calculated by Fregoso {\it et al.} in Ref. \cite{fregoso2009} as
\bea
f^{3D}_{ll^\prime;m} &=& d^2 \Big(a_{lm}^{(1)}\delta_{l,l'}
+a_{lm}^{(2)}\delta_{l,l'-2}+a_{l'm}^{(2)}\delta_{l',l-2} \Big)
\label{eq:parameter_3D}
\eea
where
\bea
a_{lm}^{(1)}&=&  \frac{4\pi (l^2+l-3m^2) (2l+1)}{l (l+1)(2l+3)(2l-1)},
\,  \nn \\
a_{lm}^{(2)} &=& -\frac{2\pi }{(l+1)(l+2)(2l+3)} \nn \\
&\times& \sqrt{[(l+1)^2-m^2][(l+2)^2-m^2]},
\label{eq:fradkin}
\eea
except for the $l=l'=m=0$ channel, where we have
$f_{00;0}^{3D}(\vec q)=V_{3D}({\vec q})$.
Please note that in Eq. \ref{eq:landauhmncs}, we use the
standard normalization convention in Ref. \cite{leggett1975,baym1991}
which is different from that in Ref. \cite{fregoso2009},
thus the parameters in Eq. \ref{eq:fradkin} are modified accordingly.
Sign errors in the original expressions of Ref. \cite{fregoso2009}
are corrected here.
It can be proved that for each $l$, $f^{3D}_{ll^\prime;m}$'s satisfy
the relation that
\bea
\sum_{m} f_{l l^\prime=l;m}^{3D}=0.
\eea

We define the average effective radial mass as
\bea
\bar m^*_{3D} =\frac{1}{4\pi}\int d\Omega_k  m^*_{3D,\pp} (\Omega_k).
\label{eq:avgmass}
\eea
The perturbative result in Eq. \ref{eq:radmass} correct to the
linear order of $\lambda$ shows $\bar m^*=m$.
Following the standard method to make the Landau matrix dimensionless,
we multiply the single component DOS.
$F^{3D}_{ll^\prime;m}=
\frac{\bar m^*}{m}N_0^{3D} f^{3D}_{ll^\prime;m}$ where $N_0^{3D}= (mk_{f0}^{3D})
/(2\hbar \pi^2)$ is the DOS of free Fermi gas.

To gain some intuition, we present some values of the low order
Landau matrix elements in terms of $\lambda$ as
\bea
F_{00;0}^{3D}(\Omega_{\vec q})&=& 4\pi \lambda P_2(\cos \theta_q),
~ F_{02;0}^{3D}=-\pi  \lambda; \nn \\
F_{11;0}^{3D}&=&\frac{18\pi}{5} \lambda;~
F_{11; \pm1 }=-\frac{9\pi}{5} \lambda; \nn \\
F_{22;0}^{3D}&=&\frac{10\pi}{7}\lambda;~
F_{22;\pm1}^{3D}=\frac{5\pi}{7}\lambda;~
F^{3D}_{22;\pm2}=-\frac{10\pi}{7}\lambda. \nn \\
F^{3D}_{13;0}&=&-\frac{3\pi}{5}\lambda,
F^{3D}_{13;\pm 1}=-\frac{\sqrt 6 \pi}{5}\lambda,\nn \\
F^{3D}_{33;0}&=& \frac{14\pi}{15}\lambda,
F^{3D}_{33;\pm 1}= \frac{7\pi}{10}\lambda,
F^{3D}_{33;\pm 2}= 0, \nn \\
F^{3D}_{33;\pm 3}&=& -\frac{7\pi}{6}\lambda,
\eea

\subsubsection{2D Landau interactions}
Similarly, in the 2D system, we define the angular distribution
$\delta n(\phi_{\vec k})$ and its decomposition in the basis of
azimuthal harmonics $e^{im\phi_{\vec k}}$ as
\bea
\delta n(\phi_{\vec k})&=&\int \frac{k dk}{(2\pi)^2} \delta n_{\vec k}
=\sum_m e^{i m \phi_{\vec k}}~ \delta n_m.
\label{eq:delta_n_2D}
\eea
The Landau interaction can be represented by a matrix as
\bea
f^{2D}(\vec k,\vec k^\prime)&=&\sum_{m,m^\prime} f_{mm^\prime}^{2D}
e^{-i m \phi_{\vec k}}
e^{i m^\prime \phi_{\vec k^\prime}},
\eea
where $f_{mm^\prime}^{2D}$ is non-zero when $m^\prime=m,m\pm 2$.

We further present our calculation for $f_{mm^\prime}$ in 2D which reads
\bea
&&f_{mm^\prime}^{2D}(\theta_0)= k_{f0}^{2D} d^2 \Big\{ P_2(\cos\theta_0) b_{m}^{(1)}
\delta_{m,m'} \nn \\
&& \ \ \ \ \ \ \ \ \ \ + \sin^2\theta_0 \Big(b_{m}^{(2)}\delta_{m,m'-2}
+b_{m'}^{(2)}\delta_{m',m-2} \Big) \Big \}, \ \ \
\label{eq:parameter_2D}
\eea
where
\bea
b_{m}^{(1)}=-\frac{8}{(2m-1)(2m+1)}, \nn \\
b_{m}^{(2)}=-\frac{2}{(2m+1)(2m+3)}.
\eea
At $\theta_0=\bar\theta_0$ where $P_2(\cos \bar\theta_0)=0$,
the dipolar interaction is purely
anisotropic (see Eq. \ref{eq:interaction_2D_k_theta_0}) and the
diagonal matrix elements  vanish.
At this particular angle, the interaction has angular
momentum $m=2$, and the matrix element is non-zero only when $m'=m\pm2$
as a result of the Wigner-Eckart theorem.

Similarly, for 2D, we also define the average radial effective mass
$\bar m^*$ which equals to $m$ at the first order of $\lambda$.
After multiplying the 2D DOS of the single component Fermi gas,
the Landau matrix becomes dimensionless $F^{2D}_{mm^\prime}= \frac{\bar m^*}{m}
N_0^{2D} f^{2D}_{mm^\prime}$ where $N_0^{2D}=m/(2\pi \hbar^2)$
is the DOS of 2D free Fermi gas.
Some low order Landau matrix elements are presented
in terms of $\lambda$ at the linear order as
\bea
F_{00}^{2D} &=& 12 \pi P_2(\cos\theta_0)\lambda; ~~
F_{11}^{2D} = -4\pi P_2(\cos\theta_0) \lambda; \nn \\
F_{22}^{2D} &=& -\frac{4 \pi}{5} P_2(\cos \theta_0)\lambda; \nn \\
F_{02}^{2D}&=&-\pi \sin^2\theta_0 \lambda;  ~~
F_{1,-1}^{2D}= 3 \pi\sin^2 \theta_0 \lambda.
\eea

Eq. \ref{eq:parameter_3D} and \ref{eq:parameter_2D} generalize the
usual Landau parameters in isotropic Fermi liquid systems to
Landau matrices in the dipolar systems.
In fact, matrix formalism is a natural generalization as long as
anisotropy enters the system.
It is the dipolar nature that constrains our Landau matrices to be
tridiagonal as shown above.

\subsection{Landau interaction matrix for two-component dipolar Fermi
gases}

We next consider the Landau interaction matrix a two-component dipolar
Fermi gas in 3D.
The intra-component Landau interaction contains both Hartree and Fock
contributions as in Eq. \ref{eq:landauHF}.
The inter-component Landau interaction only contains Hartree
contribution.
The general Landau interaction function is decomposed into density
channel response $f_s$ and spin-channel response as $f_a$
\bea
f_{\alpha\beta,\gamma\delta}(\vec k, \vec k^\prime)
=f^s(\vec k, \vec k^\prime)\delta_{\alpha\beta}\delta_{\gamma\delta}
+f^a(\vec k, \vec k^\prime) \sigma_{\alpha\beta} \sigma_{\gamma\delta},
\eea
where $f_s$ and $f_a$ at the Hartree-Fock level are expressed as 
\bea
f^s(\vec k,\vec k^\prime;\vec q)&=& V(\vec q)-
\frac{1}{2} V(\vec k-\vec k^\prime); \nn \\
f^a(\vec k,\vec k^\prime;\vec q)&=& -\frac{1}{2} V(\vec k-\vec k^\prime).
\label{eq:landautwocp}
\eea
Because the DOS at the Fermi surface for the two-component Fermi gases
is doubled compared to that of the single-component gases, the
Landau matrix elements in the density channel $F^{3D,s}_{ll^\prime;m}$
and in the spin channel $F^{3D,a}_{ll^\prime;m}$ at the Hartree-Fock level
equal to those $F^{3D}_{ll^\prime;m}$ defined for the single component case
\bea
F^{3D,s}_{ll^\prime;m}&=&F^{3D,a}_{ll^\prime;m}=F^{3D}_{ll^\prime;m}, \nn \\
\eea
if at least one of $l$ and $l^\prime$ are nonzero.
The case of $l=l^\prime=0$ is special, for 3D
we have
\bea
F^s_{00;0}=2 F_{00;0}; \ \ \ F^a_{00;0}=0.
\eea

In 2D, a similar conclusion applies at the Hartree-Fock level as
\bea
F^{2D,s}_{mm^\prime}= F^{2D,a}_{mm^\prime}=F^{2D}_{mm^\prime}, 
\eea
if at least one of $m$ and $m^\prime$ are nonzero.
For $m=m^\prime=0$, we have
\bea
F^{2D, s,a}_{00}=F^{2D}_{00}\pm \lambda
\frac{3\pi^2}{\epsilon k_f^{2D,0}} P_2(\cos\theta_0),
\eea
where $\epsilon$ is the short range cutoff defined
in Sect. \ref{sect:interaction}.

\subsection{Landau interaction matrix for $N$-component dipolar Fermi
gases}
The general $N$-component case is essentially similar in which $N$ arises
from the hyperfine multiplets.
The $SU(N)$ symmetry is very accurate since the electronic dipolar
interaction is independent of the internal hyperfine components.
A Fermi liquid theory for the 4-component Fermi gas with $SU(4)$ and
$Sp(4)$ symmetry has been constructed by one of us in Ref.
\cite{wu2003,wu2006}, which can be easily generalized to the
$N$-component case here.
For the convenience of presentation, we first define our convention
of the $N^2-1$ generators of the $SU(N)$ group
\bea
&&[M^{(1)}_{ij}]_{lk}= \delta_{il}\delta_{jk} +\delta_{ik}\delta_{jl}
~~ (1\le i < j \le n),
\nn \\
&&[M^{(2)}_{ij}]_{lk}=-i (\delta_{il}\delta_{jk}-\delta_{ik} \delta_{jl})
~~ (1 \le i < j \le n),
\nn \\
&&[M^{3}_{j}]_{lk}=\frac{\mbox{diag}(1, ..., 1, -(j-1),0,...)}
{\sqrt {j(j-1)/2}} ~(2 \le j \le n), \nn \\
\eea
where $M^{1}$, $M^{2}$ and $M^{3}$ are the $SU(N)$ version of the
Pauli matrices of $\sigma_1$, $\sigma_2$ and $\sigma_3$, respectively.
Then the $SU(N)$ Fermi liquid Landau interaction can be written
as
\bea
f_{\alpha\beta,\gamma\delta}(\vec k, \vec k^\prime)
&=&f^s(\vec k, \vec k^\prime)\delta_{\alpha\beta}\delta_{\gamma\delta}
+f^a(\vec k, \vec k^\prime) \nn \\
&\times&
\Big \{ \sum_{ij} (M^{1}_{ij;\alpha\beta} M^{1}_{ij;\gamma\delta}
+ M^{2}_{ij;\alpha\beta} M^{2}_{ij;\gamma\delta}) \nn \\
&+&
\sum_i M^{3}_{i;\alpha\beta} M^{3}_{i;\gamma\delta} \Big\},
\eea
where $f_s$ and $f_a$ are expressed as
\bea
f^s(\vec k,\vec k^\prime;\vec q)&=& V(\vec q)-
\frac{1}{N} V(\vec k-\vec k^\prime); \nn \\
f^a(\vec k,\vec k^\prime;\vec q)&=& -\frac{1}{N} V(\vec k-\vec k^\prime).
\eea

Again due to the DOS at the Fermi surface is $N$-times enhanced,
the Landau matrix elements for the $N$-component dipolar gas
in the density channel $F^{3D,s}_{ll^\prime;m}$ and in the spin channel
$F^{3D,a}_{ll^\prime;m}$ at the Hartree-Fock level equal 
to those $F^{3D}_{ll^\prime;m}$ defined for the single component case as
\bea
F_{ll^\prime;m}^s=F_{ll^\prime;m}^a=F_{ll^\prime;m}
\eea
when at least one of $l$ and $l^\prime$ are nonzero.
When $l=l^\prime=0$, we have
\bea
F^s_{00;0}=N F_{00;0}; \ \ \ F^a_{00;0}=0.
\eea
In other words, $F^s_{00;0}$ has a large $N$-enhancement
compared to all of the other Landau matrix elements.

Again, in 2D, a similar conclusion applies at the Hartree-Fock level as
\bea
F^{2D,s}_{mm^\prime}= F^{2D,a}_{mm^\prime}=F^{2D}_{mm^\prime}, 
\eea
if at least one of $m$ and $m^\prime$ are nonzero.
For $m=m^\prime=0$, we have
\bea
F^{2D, s}_{00}&=&F^{2D}_{00}+ \lambda
(N-1)\frac{3\pi^2}{\epsilon k_f^{2D,0}} P_2(\cos\theta_0), \nn \\
F^{2D, a}_{00}&=&F^{2D}_{00}-
\lambda \frac{3\pi^2}{\epsilon k_f^{2D,0}} P_2(\cos\theta_0).
\eea

\section{Thermodynamic quantities}
\label{sect:thermo}
We next consider the renormalization to the thermodynamic
properties from Landau interaction matrices.
For the simplicity of presentation, we only consider the
single-component dipolar systems here.
With slight modifications, the results also apply
to the  multi-component cases.

\subsection{Effective mass and Landau interaction matrix elements}

In this subsection, we rederive the effective mass renormalization 
discussed in Sect. \ref{sect:anisotropic_fermi_surface} using 
Landau matrix formalism. 
We will see how the Landau parameters in Eq. \ref{eq:parameter_3D} 
and \ref{eq:parameter_2D} enter in the effective masses in 3D and 
2D respectively. 
The formalism below is general for any anisotropic Fermi 
liquid system (with azimuthal symmetry in 3D).

In Galilean invariant systems, the fermion effective mass renormalization
$m^*/m=(1+\frac{1}{3}F^s_1)$ is an important result of
the isotropic Fermi liquid theory.
This results in the same renormalization factor for the DOS
at Fermi surface and the specific heat as
$C_{FL}/C_{FG}=m^*/m$, where $C_{FL}$
and $C_{FG}$ are specific heat for Fermi liquid and ideal Fermi
gas, respectively.

For the anisotropic 3D dipolar systems with the Galilean invariance,
we present a similar result.
The relation connecting effective mass and bare mass
still holds for anisotropic interactions \cite{landau1957,landau1959}
\bea
\frac{\partial \epsilon(\vec k)}{\partial \vec k}
&=&\frac{\vec k}{m}+\int \frac{d^3 \vec k^\prime}{(2\pi)^3}
f^{3D}(\vec k; \vec k^\prime) \frac{\partial n(\epsilon (\vec k^\prime) )}
{\partial \vec k^\prime}.
\label{eq:effmass2}
\eea
However, for anisotropic systems, a self-consistent solution to
Eq \ref{eq:effmass2} has to be done numerically.
To the linear order of $\lambda$, we perform the analytic
calculation by approximating $\epsilon(\vec k^\prime)$ in
the RHS of Eq. \ref{eq:effmass2} with the free fermion energy
as follows.
We take the radial derivative of Eq. \ref{eq:effmass2},
\bea
\frac{1}{ m^*_{3D,\pp} (\theta_k)}&=&\frac{ 1}{ m} - \frac{N_0^{3D}}{m}
\int d\Omega_{k^\prime}
f^{3D}(\vec k-\vec k^\prime) (\hat k \cdot \hat k ^\prime) \nn \\
&=&\frac{1}{m}\Big[ 1 - \tilde F_{11,\pp}(\theta_k) - \tilde F_{13,\pp}(\theta_k) \Big ],
\label{eq:effmass3}
\eea
where $m^*_{3D,\pp} (\theta_k)$ is the radial effective mass
given in Eq. \ref{eq:radmass}.
$\tilde F_{11,\pp}(\theta_k)$ and $\tilde F_{13,\pp}(\theta_k)$ in Eq. \ref{eq:effmass3}
are the angular dependent Landau parameters defined as follows
\bea
\label{eq:f11q}
\tilde F_{11,\pp}(\theta_k)&=& \frac{4\pi}{3}\sum_m \frac{F^{3D}_{11;m}}{3}
|Y_{lm} (\theta_k,0)|^2 \nn \\
&=&\frac{18\pi \lambda}{5} P_2(\cos \theta_k),\\
\tilde F_{13,\pp}(\theta_k)&=&
\frac{4\pi}{3}\sum_{m=0,\pm1} \frac {F^{3D}_{13;m}}{\sqrt {21}}
 Y^*_{3m}(\theta_k,0) Y_{1m}(\theta_k,0)\nn \\
&=& \frac{\sqrt{21}\pi}{5} \lambda P_2(\cos \theta_k).
\eea
Eq. \ref{eq:effmass3} is a generalization to that of the
isotropic case of  $m^*/m=1+\frac{F^s_1}{3}$
which can be rewritten as
\bea
\frac{1}{m^*}=\frac{1}{m} -\frac{ N_0^{3D}}{m} \frac{f_1^s}{3}.
\eea
Thus to the linear order of $\lambda$,
\bea
\frac{1}{m_{3D,\pp}^*(\theta_k)}
=\frac{1}{m} \big\{ 1-\pi\lambda P_2(\cos\theta_k) \big\},
\eea
which agrees with Eq. \ref{eq:radmass}.
At the linear order of $\lambda$, as an averaged effect,
there should be no specific heat renormalization due
to the dipolar interaction.

Parallel analysis can be carried out for the 3D transverse
effective mass as
\bea
\frac{1}{m^{*}_{3D,\perp}(\theta_k)}&=&\frac{1}{m}
\big[\tilde F_{11,\perp}(\theta_k)+\tilde F_{13,\perp}(\theta_k)],
\eea
where
\bea
\tilde F_{11,\perp}(\theta_k)&=& \frac{1}{6}
\big (F_{11;0}^{3D}-\frac{F_{11;1}^{3D}+F_{11;-1}^{3D}}{2} \big)
\sin2\theta_k,\nn \\
\tilde F_{13,\perp}(\theta_k)&=& -\frac{4\pi}{3} \sqrt{\frac{2}{21}}
\Big [F_{13;0}^{3D}Y_{11}(\theta_k,0)Y_{30}(\theta_k,0) \nn \\
&+&\frac{Y_{10}(\theta_k,0)}{2}   \Big(Y_{3,-1}(\theta_k,0) F_{13;-1}^{3D} \nn \\
&-&Y_{31}(\theta_k,0)  F_{13;1}^{3D}\Big) \Big].
\eea
For the 2D effective mass, the Landau parameters renormalizations are summarized below:
\bea
\frac{1}{m^{*}_{2D,\pp}(\phi_k,\theta_0)}&=&\frac{1}{m}
\big[1-F_{1,1}^{2D}-(F^{2D}_{1,-1}+F^{2D}_{3,1})\cos2\phi_k\big], \nn \\
\frac{1}{m^{*}_{2D,\perp}(\phi_k,\theta_0)}&=&\frac{1}{m}
\big (F^{2D}_{3,1}-F^{2D}_{1,-1}\big)\sin2\phi_k,
\eea
They all agree with the previous results in Sect. \ref{sect:anisotropic_fermi_surface} based on Hartree-Fock calculation, just as expected.

\subsection{Thermodynamic susceptibilities}
\label{subsect:susp}

We study the variation of the ground state energy of the dipolar
Fermi gas respect to Fermi surface distortion.
We define the variation of the angular distribution is defined as
\bea
\delta n(\Omega_k) =\int d k \frac{k^2 }{(2\pi)^3} \delta n(k,\Omega_k),
\label{eq:angle}
\eea
where only the radial integral of $k$ is performed.
The total density variation can be expressed as
$\delta n=\int d\Omega_k \delta n(\Omega_k)$.
The spherical harmonic expansion of $\delta n(\Omega_k)$ is defined
as
\bea
\delta n (\Omega_k) =\sum_{lm} \delta n_{lm} Y_{lm} (\Omega_k).
\label{eq:harmonics}
\eea

The corresponding ground state energy is represented as
\bea
\frac{\delta E}{V}&=&\delta E_{kin} +\delta E_{int}-
4\pi h^{ex}_{lm} \delta n_{lm},
\eea
where the first two terms are the variation of kinetic and interaction
energies respectively, and the last term is the coupling to the
external fields with partial wave channels of $lm$.
Expanding the Hartree-Fock single particle energy around $k_{f_0}^{3D}$ as
\bea
\epsilon^{HF}_{3D}(k, \Omega_k)&=& \epsilon_0 (k_{f0}^{3D})+
\Sigma^{HF}_{3D}(k_{f0}^{3D},\Omega_k) \nn \\
&+& \frac{\hbar k_{f_0}^{3D}}{m^*_{3D,\pp}(\Omega_k)} (k-k_{f0}^{3D}).
\eea
The variation of the kinetic energy is represented as
\bea
\frac{\delta E_{kin}}{V}&=& \int d\Omega_k \Big\{ 2\pi
[\frac{m^*_{3D,\pp} (\Omega_k)}{m} N_0^{3D}]^{-1} [\delta n(\Omega_k)]^2
\nn \\
&+&\Sigma_{3D}^{HF}(k_{f_0}^{3D}, \Omega_k) \delta n(\Omega_k) \Big\},
\label{eq:kinetic1}
\eea
Eq. \ref{eq:kinetic1} can be expressed as
\bea
\frac{\delta E_{kin}}{V}&=& 2\pi  (N_0^{3D} \frac{\bar m^*}{m})^{-1}
\sum_{ll^\prime m}   \delta n^*_{l^\prime m} M_{ll^\prime;m}^{3D} \delta n_{lm}\nn \\
&-&4\pi h_{20}^0 \delta n_{20},
\eea
where $h_{20}^0=\frac{2}{3}\sqrt{\frac{\pi}{5}}\lambda E_{k_{f0}}^{3D}$,
and $\bar m^*$ is defined in Eq. \ref{eq:avgmass}.
The perturbative results at the linear order of $\lambda$ give rise to
\bea
M_{ll^\prime;m}^{3D}= m_{lm}^{(1)} \delta_{ll^\prime}+  m_{lm}^{(2)} \delta_{l, l^\prime-2}
+ m_{lm}^{(2)} \delta_{l,l^\prime-2},
\eea
where
\bea
m_{lm}^{(1)}&=&1+\pi  \frac{ ( l^2+2l-3 m^2)}{(2l+3)(2l-1)}\lambda  \nn \\
            &=&1+\frac{l(l+1)}{4(2l+1)}a^{(1)}_{lm}\lambda , \\
m_{lm}^{(2)}&=&-\frac{3\pi }{2(2l+3)}\sqrt{ \frac{[(l+1)^2-m^2][(l+2)^2-m^2]}{(2l+1)(2l+5)}}\lambda\nn \\
            &=& \frac{3(l+1)(l+2)}{4\sqrt{(2l+1)(2l+5)}}a^{(2)}_{lm}\lambda.
\eea

The variation of the interaction energy reads
\bea
\frac{\delta E_{int}}{V}=\frac{1}{2}
\sum_{lm} f_{ll^\prime,m}^{3D} \delta n^*_{lm} \delta n_{lm}.
\eea

With the total field $h_{lm}= h^{ex}_{lm}+ h_{20}^0 $,
the ground state energy is represented as
\bea
\frac{\delta E}{V}&=&\delta E_{kin} +\delta E_{int} -4\pi h_{lm} \delta n_{lm}
\nn \\
&=& 4\pi \Big\{ \frac{1}{2\chi_0}\sum_{ll^\prime;m}\delta n_{lm}^* \delta
n_{l^\prime m}
K_{ll^\prime;m}^{3D}-h_{lm} \delta n_{lm} \Big\}, \ \ \ \ \ \
\eea
where $\chi_0=\frac{\bar m^*}{m}N_0^{3D}$.
The matrix kernel $K_{ll^\prime}$ reads
\bea
K_{ll^\prime;m}^{3D}= M_{ll^\prime;m}^{3D}
+\frac{F_{ll^\prime;m}^{3D}}{\sqrt{(2l+1)(2l^\prime+1)}}.
\label{eq:m_matrix_3D}
\eea
The the expectation value of the $\delta n_{lm}$ in the field of $h_{lm}$
can be straightforwardly calculated as
\bea
\delta n_{lm} =\chi_0  (K^{3D}_{ll^\prime;m})^{-1} h_{l^\prime m}.
\label{eq:renormalized_suscept}
\eea
Thus $\chi_0 (K^{3D}_{ll^\prime;m})^{-1}$ is the renormalized susceptibility
matrix for a 3D dipolar Fermi system.

A parallel study can be applied to 2D, where we replace the
quasiparticle density fluctuation $\delta n_{lm}$ by
$\delta n_{m}$ as defined in Eq. \ref{eq:delta_n_2D}.
The corresponding result is similar to the 3D case,
except we replace the $M^{3D}_{ll'm}$ matrix by
\bea
K_{mm'}^{2D}=M_{mm'}^{2D}+F_{mm'}^{2D}.
\label{eq:m_matrix_2D}
\eea
where
\bea
M^{2D}_{mm'}&=&\Big[1+4 \pi \lambda P_2(\cos\theta_0) \Big]\delta_{mm'} \nn \\
 &&\ \ \ \ -\frac{7\pi}{5}\lambda\sin^2\theta_0 \big(\delta_{m,m'-2}+\delta_{m',m-2}  \big).
\eea

The renormalized susceptibility in 2D has the same form
as in Eq. \ref{eq:renormalized_suscept}.

\subsection{Thermodynamic stability}
\label{subsect:therm}
In isotropic Fermi liquids, Fermi surface becomes unstable if anyone of
the Landau interaction parameters $F_l$ is negatively large enough,
i.e., $F_l <-(2l+1)$.
This can be understood by treating Fermi surface as elastic membrane.
The kinetic energy always contributes to positive surface tension,
while interaction contributions can be either positive or negative
depending of the sign of $F_l$ in each channel.
If the negative contribution from interaction exceeds the kinetic
energy cost, Fermi surface distortion occurs.

In the 3D anisotropic dipolar Fermi gas, we diagonalize the interaction
matrix $K_{ll^\prime;m}^{3D}$ as
\bea
K_{ll^\prime;m}^{3D} = T_{m}^{-1} \mbox {diag}\{\mu_{0}^m,
\mu_{1}^m, \mu_{2}^m, ...  \} T_m.
\eea
The thermodynamic stability conditions can be similarly stated as each
of $\mu^m_i$ is positive, {\it i.e.},
\bea
\mu^m_i>0
\label{eq:instability_3D}
\eea
for arbitrary $m$ and $i$.
It is not difficult to observe that in 2D, we simply need to replace
the matrix by $K_{mm'}^{2D}$ defined in Eq. \ref{eq:m_matrix_2D}.
In the isotropic systems, $K_{ll^\prime;m}^{3D}$ becomes diagonal,
and this stability criterion reduces back to that of the Pomeranchuk.

\begin{figure}[tb]
\centering\epsfig{file=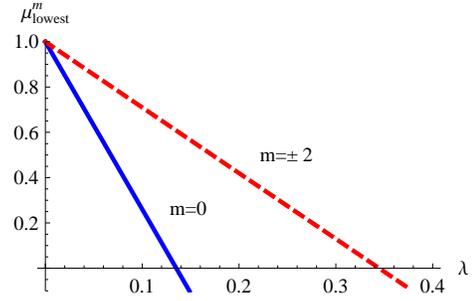,clip=1,
width=0.7\linewidth,angle=0}
\caption{A plot of the lowest eigenvalues $\mu^m_i$ of $K_{ll';m}^{3D}$
versus $\lambda$ for $m=0,\pm 2$. 
The instabilities of the $s$-channel (solid blue)
and $d_{\pm 2}$-channel (dashed red) are revealed by the vanishing
of their eigenvalues when $\lambda$ passes 0.135 and 0.35 respectively.}
\label{fig:instability_3D}
\end{figure}

For our anisotropic dipolar system, we diagonalize the $K$-matrix
numerically and determine the instability conditions.
For example, we compare two strongest instabilities in the sectors
of $m=0$ and that of $m=\pm 2$ with even $l$.
The former mainly lies in the $s$-channel and the latter mainly
lies in the $d_{x^2-y^2\pm 2 ixy}$-channel,  both of which hybridize
with other even partial wave channels with the same values of
$m=0,\pm 2$, respectively.
Their eigenvalues are denoted as $\mu_{s}$ and $\mu_{d_{\pm 2}}$, respectively.
The $F^{3D}_{00;0}(\vec q)$ explicitly depends on the orientation of $\vec q$.
We put $\vec q$ along the equator which results in the minimal eigenvalues
of $K_{ll;m}^{3D}$, and plot $\mu_s$ and $\mu_{d_{\pm2}}$ in Fig. \ref{fig:instability_3D}
The $s$-channel eigenvalue
becomes zero at $\lambda_s=0.135$, and the $d_{x^2-y^2\pm 2i xy}$-channel
eigenvalue becomes zero at  $\lambda_{d_{\pm2}}=0.35$.
The $s$-channel instability corresponds to the Fermi surface collapse
perpendicular to the dipole orientation,
and the $d$-channel one is the biaxial nematic instability of the Fermi surface
as studied in Ref. \cite{fregoso2009} (If we truncate the matrix
$K^{3D}_{ll';0}$ at $l=4$, $\lambda_{d_{\pm2}}$ will become 0.95,
which is essentially the finding in Ref. \cite{fregoso2009}.
Therefore, the $d$-channel instability actually occurs earlier
than what they expect in Ref. \cite{fregoso2009})
The $s$-channel instability occurs before the $d$-channel
instability with the purely dipolar interaction.
Nevertheless, the $s$-channel instability can be cured by introducing
a positive non-dipolar short-range $s$-wave scattering potential $V_{00;0}$,
which adds to the Landau parameter of $F_{00;0}^{3D}$ without affecting
other channels.

Compared to the compressibility calculation by the variational
method in Ref. \cite{sogo2008}, in which it shows a stability
condition equivalent to $\lambda\approx 0.42$.
This is larger than our criterion $\lambda_s=0.135$ mentioned above.
However, the calculation in Ref. \cite{sogo2008} did not take
into account the Hartree contribution to the ground state energy.
Although it is zero for rigorously homogeneous systems,
it is actually singular due to the singularity
of the Fourier transform of the 3D dipolar interaction $V_{3D}(\vec q)$
as $\vec q\rightarrow 0$.
Assuming an infinitesimal density wave vector $\vec q$ in the ground
state, the Hartree self-energy contribution appears
in $F_{00;0}(\vec q)^{3D}$, and becomes most negative
at $\theta_q=\frac{\pi}{2}$.
In other words, Ref. \cite{sogo2008} sets $F^{3D}_{00;0}(\vec q)=0$
which overestimates the stability of the 3D dipolar gas.
Our result is supported by the numerical calculation given
in Ref. \cite{ronen2009}, in which the Fermi surface instability
manifests as the onset of an unstable collective mode
at $\lambda_s\sim0.14$.
For the multi-component case, the critical value should be
further suppressed by a factor of $1/N$ because DOS is $N$
times large, where $N$ is the number of components.

Similar static instability also occurs in the 2D dipolar system.
Let us consider the specific case $\theta_0=\bar\theta_0$,
which corresponds to the most anisotropic dipolar gas in 2D
(Eq. \ref{eq:interaction_2D_k_theta_0}).
We numerically diagonalize the $K_{mm'}^{2D}$ matrix in
Eq. \ref{eq:m_matrix_2D}.
The result shows that an instability would occur at $\lambda\sim 0.15$.


\section{Collective excitations in the density and spin channels
in 3D}
\label{sect:collective mode}

In this section, we study the density-density response of the dipolar
Fermi liquid and the corresponding collective excitations.
The spin channel collective modes will also be studied in multi-component
systems.
Due to the anisotropic nature of the dipolar interaction, the response
function exhibits anisotropic features.
It shows the collective excitation of the zero sound
which only propagates within a certain range of directions with
anisotropic dispersion relations but become damped in other directions.

In the following, we will first present the generalized dynamical response
of the dipolar system for 2 and 3D in Sect. \ref{subsect:generalized_response},
which is followed by the 3D collective excitations in
Sect. \ref{subsect:zero_3D}.
In order to obtain a clearer picture about the contribution of each mode
towards the zero sound, we first consider the simplest $s$-wave channel
in 3D only, where we will see how the zero sound propagation is
restricted in certain propagation direction relative to the external
electric field orientation.
We then proceed to a more quantitative calculation by considering
the correction from the $p$-channel. After that, we discuss the
possible spin collective spin mode in Sect. \ref{subsect:zero_spin}.

\subsection{Generalized dynamical response functions} 
\label{subsect:generalized_response}

For the purpose to study collective modes, we define $\delta\nu_{\vec p}$
as the variation of Fermi distribution respect to the equilibrium
Fermi surface with the dipolar interaction strength $\lambda$.
Compared to the definition $\delta n_{\vec p}$ in Eq. \ref{eq:FSvariation}
and $\delta n (\Omega_k)$ in Eq. \ref{eq:angle}
which refer to the non-interacting Fermi surface,
their spheric harmonic components are the same except
$\delta \nu_{20}=\delta n_{20}-\delta n^0_{20}$ in the channel of
$(l,m)=(2,0)$.
$n^0_{20}$ refers to the equilibrium Fermi surface nematic
distortion.

To start with, we consider the standard Boltzmann transport equation
for the Fermi liquid \cite{baym1991,negele1988}
\bea
\delta \nu_{\vec p}+\frac{\vec v_{\vec p}^{3D} \cdot \vec q}{\omega - \vec v_{\vec p}^{3D}
\cdot \vec q}
\frac{\partial n_p}{\partial \epsilon_{HF,p}}
\int \frac{d^3p'}{(2 \pi)^3} f^{3D}(\vec p,\vec p') \delta \nu_{\vec p'}=0.
 \ \ \
\eea
where  $\vec v_{\vec p}^{3D}$ is the Fermi velocity in Eq. \ref{eq:fermivelocity_3D},
$\epsilon_{HF,p}$ is the Hartree-Fock single particle spectrum and
$f^{3D}(\vec p, \vec p')$ being the 3D Landau interaction
in Eq. \ref{eq:landauHF}.
This equation is equivalent to
\bea
\delta \nu(\Omega_p)-   \frac{ N^{3D}(\Omega_p) \vec v_{\vec p} \cdot \vec q}
{\omega - \vec v_{\vec p} \cdot \vec q}
\int \frac{d\Omega_{p^\prime}}{4\pi} f_{\vec p \vec p^\prime} 
\delta \vec \nu(\Omega_{p^\prime}) =0.
\eea

After the spherical harmonic decomposition, we arrive at the generalized
transport equation:
\bea
\sum_{l'm'} \Big\{\delta_{ll'}\delta_{mm'}+ \sum_{l^{\prime\prime}}
\chi_{ll^{\prime\prime};mm'}^{3D}(\omega, \vec q)
F^{3D}_{l^{\prime\prime}l';m'}   \Big\}\delta  \nu_{l'm'}=0, \ \ \
\label{eq:boltzmann_3D}
\eea
where
\bea
\chi_{ll';mm'}^{3D}(\omega,\vec q) &=& -\frac{1}{\sqrt {(2l+1)(2l'+1)}}
\int d \Omega_p \frac{N^{3D}(\Omega_{\vec p})}{N^{3D}_0} \nn \\
&\times& Y^*_{lm}(\Omega_{\vec p})
\frac{\vec v_p^{3D} . \vec q}{\omega - \vec v_p^{3D} . \vec q}
Y_{l'm'}(\Omega_{\vec p}),
\label{eq:response_3D}
\eea
where $F^{3D}_{ll';m}$ is the Landau parameters defined
in Eq. \ref{eq:parameter_3D}.
Due to the dipolar anisotropy and the propagation direction of $\vec q$,
$\delta \nu_{lm}$ in different channels are coupled.
The dispersion of the collective modes can be obtained by equating
the determinant of the above matrix equation to zero.
The formalism above is not restricted to dipolar system and can by
applied to any 3D Fermi liquid with azimuthal but not rotational symmetry.

Generally speaking, in order to calculate the zero sound propagation
of these anisotropic systems, one has to deal with the above infinite
matrix equation. However, in this 3D dipolar system, it turns out that
the physics of zero sound is well captured by considering only the $s$
and longitudinal $p$ channels.
In order to avoid complication, in the following, we start by
examining the $s$-channel only, where the anisotropic feature of
zero sound already appears in terms of a limited propagation direction
and an anisotropic sound velocity.
Afterwards, we consider the effect of the longitudinal $p$-wave mode,
where a quantitative zero sound velocity is obtained as a function
of propagation angle and this is in good agreement with a numerical
study performed in Ref.\cite{ronen2009}.


\subsection{The 3D density channel collective mode: zero sound}
\label{subsect:zero_3D}

\subsubsection{The $s$-wave channel}
\label{subsect:zero_swave}

Let us warm up by only keeping the $s$-wave channel component in Eq.
\ref {eq:response_3D}.
To the lowest order of $\lambda$, we only keep the anisotropic
Landau parameter of $F_{00;0}(\vec q)$ which explicitly depends
on the direction of $\vec q$, and completely neglect the anisotropic
DOS and Fermi velocity.
Thus the $\chi_{00;00}^{3D}(\omega,\vec q)$ is simply the standard
textbook result which reads at small values of $s=\omega/v_{f0}^{3D} q\ll 1$
as
\begin{eqnarray}
\chi_{00;00}^{3D}(\vec{q},\omega) =N^{3D}_0 \Big\{1-{s\over 2 }
\ln |{1+s \over {1-s}}|  + i {\pi \over 2}s\Theta(s<1) \Big\}. \nn \\
\label{eq:bare response1}
\end{eqnarray}
The collective excitation is determined by the pole of it:
\bea
1+ F^{3D}_{00;0}(\Omega_{\vec q}) \chi^{3D}_{00;00}=0.
\label{eq:zero sound 3D_swave}
\eea
$s>1$ is needed to ensure the collective mode underdamped.

Eq. \ref{eq:zero sound 3D_swave} gives rise to an anisotropic zero
sound velocities which explicitly depends on the propagation
direction of $\theta_q$ as
\begin{eqnarray}
\omega^{(0)}_{3D}(q, \theta_q)=c_{3D}^{(0)}(\theta_q)q,
\end{eqnarray}
where
\begin{eqnarray}
\frac{c_{3D}^{(0)}(\theta_q)}{v_{f0}^{3D}}
=\left\{ \begin{array}{rr}
1+2e^{-\frac{1}{4\pi \lambda P_2(\cos\theta_q)}} ;
&\mbox{ $ 8\pi\lambda P_2(\cos\theta_q) \ll 1$} \\
\sqrt{\frac{8\pi}{3} \lambda P_2(\cos\theta_q) };
&\mbox{$ 8\pi\lambda P_2(\cos\theta_q) \gg 1$}
\end{array} \right. \nn \\
\label{eq:zero sound 3D}
\end{eqnarray}
for $\theta_q< \bar{\theta}_0$  or $ \theta_q>\pi- \bar{\theta}_0$,
where $\bar \theta_0$ is defined in Eq. \ref{eq:thetabar}.
The zero sound is well-defined around the north and
south poles of the Fermi surface, where the interaction is most repulsive.
On the other hand, for $\bar\theta_0<\theta_q<\pi-\bar\theta_0$,
$F_{00;0}(\Omega_q)$ becomes negative, the solution shows $s<1$
which means the dispersion goes into the particle-hone continuum
and ceases to be a sharp collective mode.

\begin{figure}[tb]
\centering\epsfig{file=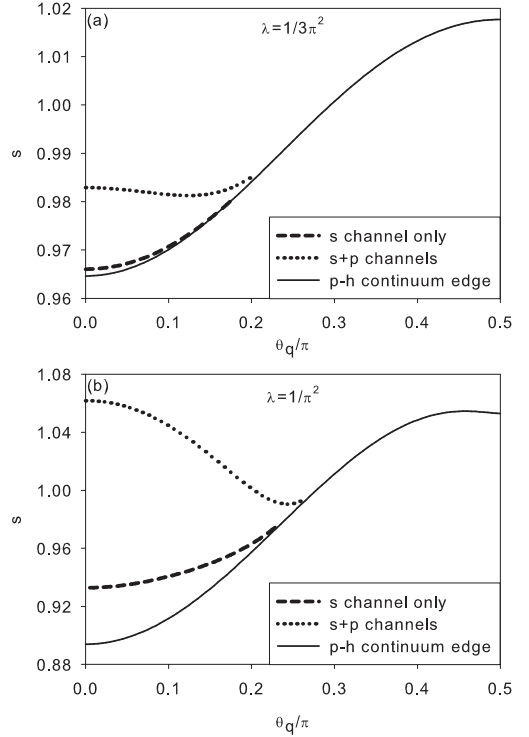,width=0.8\linewidth,angle=0}
\caption{Dispersions of the zero sound $s(\theta_q)=\omega
(\theta_q)/v_{f0}^{3D} q$ for the pure dipolar interaction
including the coupling between the $s$-wave
and the longitudinal $p$-wave modes for $\lambda = \frac{1}{3\pi^2},
 \frac{1}{\pi^2}$.
The $s$-wave approximation is included for comparison
which is valid in the limit of a large component number $N$
with the replacement of $\lambda\rightarrow N\lambda$.
When the sound speed hits the particle hole continuum,
the sound becomes damped.
These results are in good agreement with a numerical study
in Ref. \cite{ronen2009}.
(The parameters are chosen for a better comparison with
 the Ref. \cite{ronen2009}, where their parameters
$D=3\pi^2\lambda$ are chosen to be 1 and 3.)
}
\label{fig:zero_3D_pwave}
\end{figure}

In order to refine the above calculation, one has to take into account
the anisotropic Hartree-Fock single particle spectra and thus the
anisotropic Fermi surface.
In this sense, the full $\chi^{3D}_{00;00}(\omega, \vec q)$ function
is given by
\bea
\chi_{00;00}^{3D}(\omega,\vec q)&=&-
\int \frac{d\Omega_p}{4\pi N^{3D}_0}
\frac{ N^{3D}(\Omega_p)(\vec q \cdot \nabla_p \epsilon_{HF}(\vec p))}{
\omega-\vec q \cdot \nabla_p \epsilon_{HF}(\vec p)+i\eta}\nn \\
&=& 1 -\int \frac{d\Omega_p}{4\pi}
\frac{N^{3D}(\Omega_p)}{N^{3D}_0}\frac{s}{s-f(\Omega_p, \Omega_q)+i\eta}, \nn \\
\eea
where $N^{3D}(\Omega_p)=N_0 [1+\frac{5\pi}{3}P_2(\cos\theta_p)]$
defined in Eq. \ref{eq:3DDOS} and $\epsilon_{HF}(\vec k)$ is the
Hartree-Fock single particle spectra. The angular form factor
$f(\Omega_p,\Omega_q)$ is defined as
\bea
f(\Omega_p, \Omega_q)&=& \frac{\vec q \cdot \nabla_p \epsilon_{HF}(\vec p)}
{v_{f0}^{3D} q} = [1-\frac{\pi}{3}\lambda P_2(\cos\theta_p)]\nn \\
&\times& \big[\sin \theta_q \sin\theta_p \cos \phi_p
+ \cos\theta_q \cos \theta_p \big]
\nn \\
&+& \lambda \pi \sin 2\theta_p
(\sin\theta_q \cos\theta_p \cos\phi_p \nn \\
&-&\cos\theta_q \sin\theta_p),
\eea
where the propagation direction $\vec q$ is chosen in the $xz$-plane
with the polar angle $\theta_q$. Eq. \ref{eq:zero sound 3D_swave}
is solved numerically and plotted in Fig. \ref{fig:zero_3D_pwave}
along with the edge of particle-hole continuum and the further
refined result including the $p$-wave channel correction.
(We postpone the discussion of it until we include the $p$-channel
correction in the following subsection.)

So far, we neglect the coupling between the $s$-wave and other
high partial channel  Fermi surface excitations.
This approximation is valid if the Landau matrix elements
in other channels are small compared to that in the $s$-wave channel.
This is justified in the multi-component case with a large $N$
in which $F_{00;0}$ has a large $N$-enhancement compared to all
of other matrix elements.
Thus the result of Eq. \ref{eq:zero sound 3D} is correct in the
large $N$-limit with the replacement of $\lambda$ by $N\lambda$.
However, for the single component dipolar interaction, i.e.,
$N=1$, the zero sound dispersion is modified from
the coupling to the longitudinal $p$-wave channel modes as explained
below.

\subsubsection{The correction from the coupling to the
longitudinal $p$-wave channel}

The coupling between the $s$-wave and other high partial wave channel
Fermi surface excitations, say, the $p$-wave longitudinal excitation,
can significantly change the zero sound dispersion.
This effect is particularly important if the Landau interaction in
the $F_1$-channel is not small compared to $F_0$.
For example, the zero sound velocity is measured as $s=3.6\pm 0.01$
in $^3$He at 0.28 atm, while $F_0^s=10.8$ only gives $s=2$.
The inclusion of the coupling to the $p$-wave channel $F_1^s=6.3$
gives rise to the accurate value of $s=3.6$ \cite{negele1988}.
Below we consider this coupling for the single component
dipolar Fermi gas.

Now, we consider the coupling between the $s$-wave and the longitudinal
$p$-wave channel modes in our 3D dipolar system.
The formalism developed in Subsect. \ref{subsect:generalized_response}
have 3 $p$-wave modes; $l=1 $ and $m=0, \pm 1$, which mixes the
longitudinal and transverse modes.
It turns out that it is better to reformulate the generalized response
function by putting the spherical harmonic expansion $Y_{lm}$
relative to the $\vec q$ direction.
Due to the explicit anisotropy, the longitudinal and transverse
$p$-wave components should be mixed.
Nevertheless, the transverse $p$-wave channel mode is overdamped
for small positive value of Landau parameters, thus they do not
affect the zero sound much.
We only keep the mixing between $s$-wave and the longitudinal
$p$-wave modes.

The corresponding transport equation becomes a $2\times 2$
matrix equation and the collective mode can be solved from:
\bea
\mbox{Det} [ 1 + N(\vec q, \omega) ]=0,
\label{eq:zero sound 3D_p_determinant}
\eea
where the matrix kernel of $N (\vec q,\omega)$ reads
\bea
N(\vec q, \omega)=
\left( \begin{array}{cc}
\chi_{00;00}^{3D} (\vec q, \omega) F_{00;0}^{3D} (\vec q)
& \tilde\chi_{10;00}^{3D} (\vec q, \omega)
\frac{\tilde F_{110}^{3D} (\vec q)}{3} \\
\tilde \chi_{10;00}^{3D} (\vec q,\omega) F_{00;0}^{3D}  (\vec q)
& \tilde \chi_{11;00}^{3D} (\vec q, \omega)
\frac{\tilde F_{11;0}^{3D} (\vec q)}{3}
\end{array}
\right ). \nn \\
\eea
$\tilde F_{11;0}^{3D} (\vec q)$ is the longitudinal $p$-wave Landau
parameter defined as
\bea
\tilde F_{11;0}^{3D} (\vec q)
&=&  N^{3D}_0 \int d\Omega_p d\Omega_p' f^{3D}(\vec p_F , \vec p_f')
(\hat q . \hat p)^2 \nn \\
&=&\cos^2\theta_q F_{11;m=0}^{3D}+\sin^2\theta_q F_{11;m=\mp 1}^{3D} \nn \\
&=& F_{11,m=0}^{3D} P_2(\cos \theta_q)
=\frac{18\pi}{5} P_2(\cos\theta_q). \ \ \
\eea
The relevant response functions are correspondingly modified as:
\bea
\tilde \chi^{3D}_{10;00}(\vec q, \omega)
&=&- \sqrt{3} \int \frac{d\Omega_p}{4\pi}
\frac{ N^{3D}(\Omega_p)}{N^{3D}_0}\frac{  (\hat q \cdot \hat p)
f(\Omega_p, \Omega_q)}
{s-f(\Omega_p, \Omega_q)}\nn \\
\tilde \chi^{3D}_{11;00}(\vec q, \omega)&=&- 3
\int \frac{d\Omega_p}{4\pi} \frac{ N^{3D}(\Omega_p)}{N^{3D}_0}
\frac{  (\hat q \cdot \hat p)^2 f(\Omega_p, \Omega_q)}
{s-f(\Omega_p, \Omega_q)}\nn \\
\label{eq:bubble_2}
\eea
where $\vec q$ lies in the $xz$-plane with the polar angle $\theta_q$,
and $\hat q \cdot \hat k=\sin\theta_q \sin\theta_k \cos\phi_k
+\cos\theta_q \cos\theta_k$.

Before we solve Eq. \ref{eq:zero sound 3D_p_determinant},
we can get a qualitative picture by approximating the Fermi surface
and density of state to be spherical first. Under this approximation,
in the long wavelength limit $\vec q\rightarrow 0$,
Eq. \ref{eq:zero sound 3D_p_determinant} can be analytically simplified to:
\bea
\frac{-1 }{ F_{00;0}^{3D} (\vec q)+
\frac{s^2 \tilde F_{11;0}^{3D} (\vec q)}{1+\frac{\tilde F_{11;0}^{3D}  (\vec q)}{3} } }
= 1-\frac{s}{2} \ln |\frac{1+s}{1-s}| ,
\label{eq:zerospectra}
\eea
which resembles the usual zero sound condition by considering
only $F_0$ and $F_1$ parameters in an ordinary isotropic Fermi liquid
\cite{negele1988}.
The difference is that the Landau parameters now have an explicit
$\vec q$ dependence.
Because $F_{00;0}(\vec q)$ and $F_{11;0}(\vec q)$ have the same
angular dependence, including the longitudinal $p$-wave channel
mode does not change the propagation angular regime, but
enhance the zero sound velocity.

Now, we solve Eq. \ref{eq:zero sound 3D_p_determinant} numerically
by taking into account the anisotropic fermion spectra and Fermi surface.
The zero sound propagation as a function of $\theta_q$ is plotted in
Fig. \ref{fig:zero_3D_pwave}.
The calculation that only involve the $s$-channel is also included
for comparison.
There are two features: a restricted zero sound propagation angle
and an anisotropic propagation velocity, $s$.
The first feature is due to the fact that the sound enters the
particle-hole continuum for large $\theta_q$ and is thus damped.
It is clear that, in terms of the critical angle where zero sound
terminate, the $s$-wave approximation is qualitatively justified.
However, when we consider the zero sound speed as a function of
$\theta_q$, the longitudinal $p$-wave mode modifies it considerably.
These results are in a good agreement with a fully numerical calculation
based on the same Boltzmann transport theory \cite{ronen2009}.
This indicates that the physics of collective excitation of the 3D
dipolar system is well captured by considering only the $s$ and
longitudinal $p$-channels.

The zero sound velocity is affected by both the value of Landau
parameters and the Fermi velocity as shown in Eq. \ref{eq:zero sound 3D}.
For the relatively large value of $\lambda=\frac{1}{\pi^2}$, the
Landau parameters play a more important role.
Thus the sound velocity is largest for $\vec q$ along the north
and south poles where $F_{00;0}^{3D}(\vec q)$ and $F_{11;0}^{3D}(\vec q)$
are largest.
As the propagation deviates from them, the sound velocity first
becomes softened.
As the velocity hits the upper edge of the particle-hole
continuum, the zero sound ceases to propagate.
At a small value of $\lambda=\frac{1}{3\pi^2}$, Fig.
\ref{fig:zero_3D_pwave} shows the upturn of zero sound velocity
before it hits the particle-hole continuum, which is largely
determined by the anisotropic Fermi velocity.
We also plot the eigenvector for the zero sound mode for the
solution of Eq.  \ref{eq:zero sound 3D_p_determinant}.
The $s$ and the longitudinal $p$-wave components are comparable
to each other.

\begin{figure}[tb]
\centering\epsfig{file=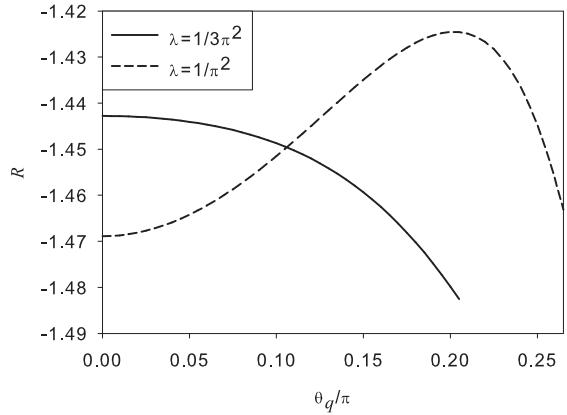,width=0.9\linewidth,angle=0}
\caption{The ratio between the longitudinal
$p$-wave and the $s$ components in the eigenvectors of the 3D zero sound, 
i.e. $R=\delta \nu_{10;0} /  \delta \nu_{00;0}$, 
for $\lambda=\frac{1}{3\pi^2}$ and $\frac{1}{\pi^2}$.
}
\label{fig:u1u0}
\end{figure}

\subsection{3D Collective spin mode in the longitudinal $p$-wave channel}
\label{subsect:zero_spin}

\begin{figure}[tb]
\centering\epsfig{file=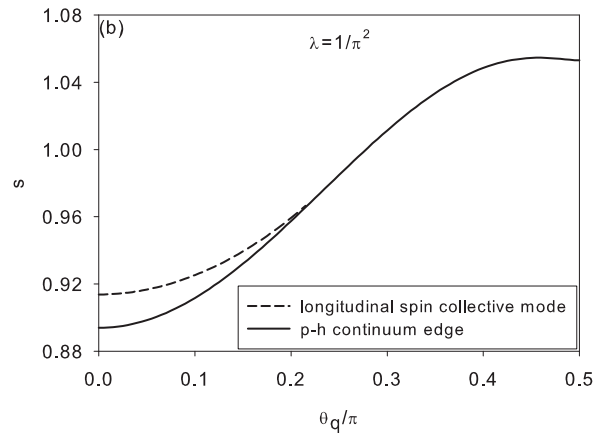,width=0.95\linewidth,angle=0}
\caption{3D collective spin mode in the longitudinal $p$-channel
for $\lambda=\frac{1}{\pi^2}$. The sound speed continuously decreases
as $\theta_q$ increases and terminates at certain critical angle.
}
\label{fig:zero_3D_spin}
\end{figure}

In addition to the zero sound mode, collective excitations
may also exist in the spin channel.
Because $F^{3D,a}_{00;0}=0$ (see definition in Eq. \ref{eq:landautwocp}),
there is no well-defined $s$-wave channel
spin excitations.
Nevertheless, the longitudinal $p$-wave channel mode becomes
a propagating mode.
The formalism to determine its dispersion is very similar to
Eq. \ref{eq:boltzmann_3D} by replacing $F^s_{ll^\prime;m}$
to $F^a_{ll^\prime;m}$.
Only keeping the longitudinal $p$-wave channel, we have
\bea
1+\tilde F^{3D,a}_{11;0}(\vec q) \tilde\chi_{11;00}^{3D}(\vec q,\omega)=0.
\eea
$\tilde F^{3D,a}_{11;0}(\vec q)=F^a_{11;0}P_2(\cos\theta_q)$ which is
the same as $\tilde F^{3D,s}_{11;0}(\vec q)$ at the Hartree-Fock level,
and $\tilde \chi^{3D}_{11;00}$ is also the same as in Eq. \ref{eq:bubble_2}.
The dispersion for this $p$-wave longitudinal spin collective mode
is plotted in Fig. \ref{fig:zero_3D_spin} for $\lambda = \frac{1}{\pi^2}$.
Again, we have an anisotropic sound velocity subject to a finite
propagation regime, where the excitation is damped when its dispersion
enters the particle-hole continuum.

In $N$-component dipolar gases with the $SU(N)$ symmetry, there
are $N^2-1$ branches of  longitudinal $p$-wave spin model.
They are essentially the fluctuating longitudinal spin current mode.
The is a novel collective mode which has not been observed
in condensed matter systems before.

\section{{Conclusions and Outlook}\label{Conclusions}}
In summary, we studied the anisotropic Fermi liquid states of the
cold atomic dipolar Fermi gases which possesses many exotic features.
The feature of the $d_{z^2-3r^2}$ symmetry of the 3D dipolar
interaction and the $d_{x^2-y^2}$ symmetry of the 2D dipolar interaction
nicely render most analytic calculations possible.
The same anisotropy exhibits at the leading order of perturbation theory
in many ways such as the Fermi surface anisotropy, Fermi velocity
and effective masses.
The Landau parameters in isotropic Fermi liquid states are generalized
to the Landau interaction matrix which has the tri-diagonal structure
as a result of the Wigner-Eckart theory.
Physical susceptibilities receive renormalization from the Landau
interaction matrix.
With large dipolar interactions, dipolar Fermi surfaces become
collapse along the directions perpendicular to the dipolar
orientation.
We also studied the collective mode in the density and spin channels.
The zero sound exhibits anisotropic dispersion relation with
largest propagation velocity along polar direction.
The $p$-wave longitudinal spin channel mode is a well defined
propagating mode for any propagation directions.

So far, we have only considered the spatial homogeneous systems.
However, the realistic experimental systems have confining traps, which 
should bring corrections to properties discussed above 
as studied in Refs \cite{sogo2008,miyakawa2008,tian2008}.
In the case of soft confining trap potentials, we can treat the
inhomogeneity by using the local density approximation.
Around each position $\vec r$, we can define a local Fermi energy 
and the local Fermi surface which are determined by the local
molecule density.
The molecule density is the highest at the center of the trap,
thus both the absolute interaction energy scale 
and the dimensionless interaction strength are strongest at the center.
Thus the thermodynamic instabilities are strongest at the center
of the trap and become weaker as moving towards the edge.
For the collective excitation of zero sound, the local sound velocity 
is largest in the center. 
As a result, the local propagation wavevector increases from
the center to the edge to maintain the excitation eigen-frequency
the same in the entire trap.
The collective modes have more server damping in the edge area
because the molecule density is low and thus is less quantum
degenerate than the center.
A detailed study of all the above effects will be presented
in a later publication.

Other open questions to be studied in the future include the issues of
higher-order (i.e. beyond Hartree-Fock theory) interaction corrections and
finite temperature corrections.  The theory developed in this work is
strictly valid for weak interaction and low temperatures.
For example, an important aspect of the dipolar Fermi gas is the finite
life time and the Fermi liquid wavefunction renormalization factor
$Z$ of quasiparticles.
Both of them naturally also develop anisotropies due to the dipolar
interaction.
However, they begin to appear at the level of the second order
perturbation theory beyond the Hartree-Fock level, thus their
anisotropies should be of an even higher order spheric harmonics
than the $d$-wave.
Another issue is the screening effect in the dipolar Fermi gases
which may change its long range negative in 3D.
These effects will be deferred to another paper
\cite{chan2009a}.

\acknowledgements
We thank helpful discussion with S. Ronen and J. L. Bohn.
C. C., C. W. and W. C. L. are supported by NSF-DMR 0804775,
ARO-W911NF0810291, and Sloan Research Foundation.
S. D. S. is supported by AFOSR-MURI and NSF-JQI-PFC.

{\it Note added} After the completion of first online version of 
the manuscript, we learned the the work by Ronen {\it et al.} 
\cite{ronen2009} in which the Fermi liquid properties of dipolar
Fermi gas are studied.



\end{document}